\documentclass[traditabstract]{aa}
\usepackage{graphicx}
\usepackage{txfonts}
\usepackage{natbib}
\bibpunct{(}{)}{;}{a}{}{,}
\defcitealias{o04}{Paper I}
\defcitealias{c08}{Paper II}
\newcommand{\fma}[1]{\mbox{$#1$}}

\newcommand{\gtsim}{\raisebox{-0.5ex}{$\;\stackrel{>}{\scriptstyle \sim}\;$}}
\newcommand{\unit}[1]{\ifmmode \:\mbox{\rm #1}\else \mbox{#1}\fi}
\newcommand{\mone}{\fma{^{-1}}}

\newcommand{\eg}{{e.g.\/}}

\newcommand{\ha}{H$\alpha$}
\newcommand{\hb}{H$\beta$}

\newcommand{\hi}{H\,{\sc i}}
\newcommand{\hii}{H\,{\sc ii}}

\newcommand{\Oi}{O\,{\sc i}}

\newcommand{\oiii}{[O\,{\sc iii}]}
\newcommand{\siii}{[S\,{\sc iii}]}

\newcommand{\feii}{Fe\,{\sc ii}}

\newcommand{\caii}{Ca\,{\sc ii}}

\newcommand{\kms}{\unit{km\,s\mone}}
\newcommand{\cm}{\unit{cm}}

\newcommand{\msun}{\unit{M$_\odot$}}
\newcommand{\msunyr}{\unit{M$_\odot$~yr\mone}}

\newcommand{\lsun}{\unit{L$_\odot$}}

\newcommand{\flux}{\unit{erg~s\mone\cm$^{-2}$}}

\newcommand{\haro}{Haro\,11}

\newcommand{\ka}{$\mathcal{A}$} 
\newcommand{\kb}{$\mathcal{B}$} 
\newcommand{\kc}{$\mathcal{C}$} 
\newcommand{\sca}{$\mathcal{CA}$} 
\def\farcs{\hbox{$.\!\!^{\prime\prime}$}}
\begin{document}
   \title{Kinematics  of Haro\,11  -- the miniature Antennae\thanks{Based on observations collected at the European
   Southern Observatory, Paranal, Chile, under observing programmes
   71.B-0602, 074.B-0771(A), 074.B-0802A }}


   \author{
          G. {\"O}stlin\inst{1} \and 
          T. Marquart\inst{2,1} \and
          R.J. Cumming\inst{3} \and
          K. Fathi\inst{1} \and
          N. Bergvall\inst{2} \and
          A. Adamo\inst{1} \and 
	  P. Amram\inst{4}  \and
	  M. Hayes\inst{1} 
          }

   \offprints{G. {\"O}stlin}

   \institute{
     Department of Astronomy, Stockholm University, SE-106~91 Stockholm, Sweden.\\
     Oskar Klein Centre, Department of Astronomy, Stockholm University, SE-106~91 Stockholm, Sweden. \\
     \email{ostlin@astro.su.se}
     \and
     Department of Physics and Astronomy, Uppsala University,
     Box 515, SE-751~20 Uppsala, Sweden
\and
     Onsala Space Observatory, Chalmers University of Technology, SE-43992 Onsala, Sweden
          \and
Aix-Marseille Universit\'e, CNRS, LAM (Laboratoire d?Astrophysique de Marseille), 13388 Marseille, France
}

   \date{Received ; accepted }
  \abstract{
   Luminous blue compact galaxies are among the most active galaxies in the local universe in  terms of their star formation rate per unit mass.
   They are rare at the current cosmic epoch, but were more abundant in the past and may be seen as the local analogs of higher redshift Lyman
   Break Galaxies. Studies of their kinematics is key to understanding what triggers their unusually active star formation. 
   In this work we investigate the kinematics of stars and ionised gas in \haro, one of the most luminous blue compact galaxies in the local universe.
   Previous works have indicated that many such galaxies may be triggered by galaxy mergers. 
   We have employed Fabry-Perot interferometry, long-slit spectroscopy and Integral Field Unit (IFU) spectroscopy to  explore the kinematics
   of \haro . We target the near infrared Calcium triplet, and use cross-correlation and penalised pixel fitting techniques to derive the stellar velocity field and
   velocity dispersion. Ionised gas is analysed through emission lines from hydrogen, \oiii , and \siii . When spectral resolution and signal to 
   noise allows we investigate the the line profile in detail and identify  multiple velocity components when present.    
   The spectra reveal a complex velocity field whose components, both stellar and gaseous, we attempt to disentangle.  
   We find that to first order, the velocity field and velocity dispersions derived from stars and ionised gas agree. Hence 
   the complexities reveal real dynamical disturbances providing further evidence for a merger in \haro . Through decomposition
  of emission lines  we find evidence 
   for kinematically distinct components, for instance a tidal arm behind the galaxy. The ionised gas 
   velocity field can be traced to large galactocentric radii, and shows significant velocity dispersion even far out in the halo. 
   If interpreted as virial motions this indicates that \haro\ may have a mass of $\sim10^{11}$\msun . Morphologically
   and kinematically \haro\ shows many resemblances with the famous Antennae galaxies, but is much denser which
   is the likely  explanation for the higher star formation efficiency in \haro . }

\authorrunning{\"Ostlin, Marquart, Cumming et al.}
\titlerunning{Kinematics of Haro11}


   \maketitle

\section{Introduction}
\label{sec:intro}

Understanding how galaxies form and evolve is one of the major goals
of contemporary astrophysics.  This involves studying the processes
which form galaxies' properties and how they change with cosmic time.  
At high redshift ($z$) our view of these processes is encumbered by
poor spatial resolution and low signal-to-noise data.
Low-$z$ analogues of the objects we see at high redshift are
therefore, if handled carefully, an indispensable complement to
direct studies of the distant universe.
 
High-$z$ galaxies selected by the Lyman-break technique
\citep[Lyman-break galaxies, LBGs;][]{Steidel99}
have proved to be 
important  for tracing the star forming galaxy population
at  $z \gtsim 3$. 
So called Lyman Break Analogues (LBAs) are low-$z$ galaxies with gross properties
(e.g. UV luminosity and surface brightness) similar to those of LBGs \citep{Heckman2005}. The number 
density of LBAs is much lower than that of high-$z$ LBGs \citep[ and references
therein]{Hoopes07}, and the closest examples   are VV\,114 \citep{Grimes06} and \haro\
\citep{Bergvall00,Hayes07} both at $z\sim 0.02$. 
 \haro\ is one of the most luminous Blue Compact Galaxies (BCGs)
known \citep[][]{KO,BO02}. Its far UV luminpsity is $L_{FUV}= 10^{10.3}L_\odot$ or  
 $0.3\, L^\star_{FUV, z=3}$ if compared
to the $z\sim3$ LBG luminosity function 
\citep[][]{Steidel99}. 

\begin{figure*}
   \centering
   \includegraphics[angle=0,width=1\textwidth]{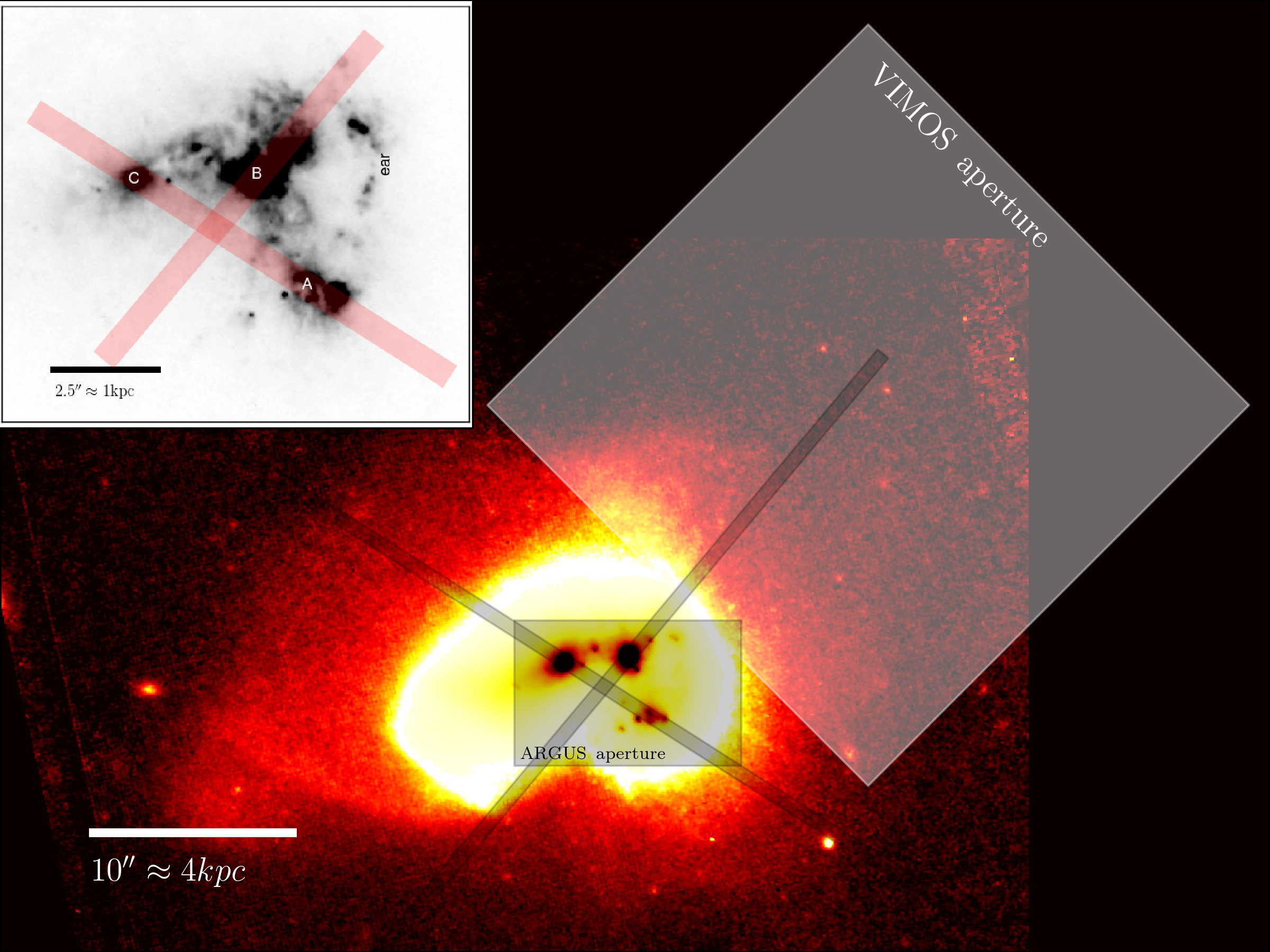}
      \caption{The colour image shows \haro\ imaged with HST/NICMOS/NIC3 in the F160W filter ($\sim$H-band). Based on a 5ks exposure obtained under general observer 
      programme 10902.
      North is up, East to the left. The scale is indicated with a solid white bar.
      The perturbed morphology is evident and demonstrates that the stellar mass distribution
      of \haro\ is very asymmetric. Telltale merger features like sharp edges and tails are clearly visible. 
      The apertures for the FLAMES/ARGUS and VIMOS/IFU spectra are indicated as grey semi-transparent rectangles.
      The greyscale image on the upper left shows the HST/ACS
     \ha\ image \citep{o09} with the three knots \ka, \kb, \kc\ and the chain like structure referred to as the 'ear' indicated. 
     The positions and orientations of the two FORS2 slits are also marked. }
         \label{nic3}
   \end{figure*}

BCGs have attracted much attention due to their properties
being similar to those expected for young galaxies in the distant
universe: low metallicity, small size, high specific star formation rate \citep{SS72}. 
BCGs are however an ill defined category, including a mixed bag of gross 
properties  \citep{Gildepaz03,Micheva12a,Micheva12b}. 
	Hence not all BCGs are extreme in terms of metallicity and specific star formation rate, but it is
among the BCGs that we find the most extreme and efficient star formation 
activities in low mass galaxies in the local universe \citep{Adamo11}.
Whereas the general triggering mechanism for starbursts in BCGs is still uncertain, there 
are indications that many of the  {\em luminous} BCGs are triggered by mergers \citep{o01,Puech,BO02,c08,Perezgallego11}.
One of the most pregnant examples is \haro , whose morphology is very irregular 
and show many features typical of merging galaxies (see Fig.   \ref{nic3}). 
  \haro\ has  three bright
starburst knots: the south-western one named \ka , the north-western one \kb \
and the eastern one \kc\ (see.~Fig.~\ref{nic3}, and also \citet{Vader93,Hayes07}). In addition,  a 
chain (which we will refer to as the ``ear'') of young massive star clusters extends
between knots \ka\ and \kb . 
Despite its small size and moderate luminosity it forms stars
at a prodigious rate \citep[$20-30$ \msun/yr,][]{Hayes07,madden2014} and with an infrared luminosity of 
$>10^{11}$\lsun\  it also qualifies as a Infrared Luminous Galaxy (LIRG).

To further put \haro\ into context, we compare it to the local universe luminosity
functions (LFs) of galaxies: \haro\ has  $L_{FUV}=4.8\,L^\star_{FUV}$ \citep[compared to $M^\star_{FUV}=-18.0$ of][]{Wyder05},
the \ha\ luminosity is $L_{H\alpha}=4.4\, L^\star_{H\alpha}$ \citep[$L^\star_{H\alpha}= 6.6\times10^{41}$erg/s/cm$^2$][]{Gallego95},
$r$-band $L_r =0.9\, L^\star_r$ \citep[$M^\star_{r}=-20.82$,][]{Blanton03} and $L_{K} =0.36\, L^\star_{K}$ \citep[$M^\star_{K}=-20.97$,][]{Kochanek01}, 
where litterature values were rescaled to the currently used H$_0$ and photometric system.
Since the $K$-band luminosity of \haro\ is still dominated by the bright central ($r<5$kpc) 
 burst component  \citep{Micheva2010}, the underlying host galaxy (contributing $\sim 30$\% 
to the $K$-band luminosity) has a stellar mass of the order of 10\% \ that of a typical quiescent $L^\star_K$ galaxy. 
  The stellar mass estimate of \citet{o01} and the observed SFR yields a specific SFR of  ($ > 10^{-9}$ yr$^{-1}$) 
or equivalently a birthrate parameter of $b > 15$.

Studies of LBAs also point to mergers as likely triggers
of the starbursts that seem to give LBGs their characteristic
properties \citep{Goncalves2010,Overzier08,Overzier09}.
%
In galaxy mergers, the detailed dynamical  interplay of dark matter, stellar components,
cold gas inflow,  and feedback from triggered star formation determine the progress of 
the starburst and how long it can carry on before being quenched, and to what extent its 
products are made available for chemical enrichment in the merger remnant itself or 
ejected to the intergalactic medium. Such processes have been modelled in large
simulations \citep[\eg][]{Nagamine04,Night06} but are still poorly resolved, and star 
formation and feedback treated through simplified recipies. Local LBG analogues like 
\haro\ have the potential to provide real observational constraints for such
simulations and improve our ability to understand galaxy evolution
over cosmic time. 

Previous kinematical studies using the \ha\ line have shown that BCGs have very  complex kinematics \citep[]{o99,o01}.
\haro\ presents multiple kinematical components  \citep[see also][]{James2013} and \citet[]{o01} found that rotation could 
not support the 
observed stellar mass.  The line width, if interpreted as virial motions, indicate a mass of 
$2\times10^{10}$ \msun \ consistent with the stellar mass.
However, given the very 
intense star formation, it is possible that the ionised gas kinematics in BCGs is driven by feedback and hence do not trace the potential \citep[e.g.][]{Green10}. 
A test of this hypothesis can be provided by observations of the kinematics of the {\em stars} in BCGs. So far results for only a
handful BCGs exist in the literature
 \citep[ESO\,400--43, He2-10, ESO\,338--04;][]{o04,m07,c08} but do not show significant differences between the 
 stellar and ionised gas velocity dispersions \citep[see also][]{kg00}. In this paper, we use a wide array of spectroscopic 
 data to analyse the kinematical status of the stellar and ionised gas components of \haro . 
We list some of \haro's basic properties in Table \ref{table:props}.

The rest of the paper is organised as follows: in Sect. 2 we present the observations and reductions from four
different ESO telescopes and instruments. In Sect. 3 we describe how the stellar kinematics was derived, and in Sect. 4 we 
do the same for the ionised gas kinematics. In sect. 5 we present the results and interpret them, including a comparison 
with the Antennae galaxies, while sect. 6 contains the conclusions.

Throughout the paper we adopt a luminosity distance of 82 Mpc ($m-M=34.6$) and a scale
of 0.4 kpc~arcsec$^{-1}$ valid for $H_0=73$ km/s/Mpc. 

\begin{table}
\caption{Basic properties of \haro}             
\label{table:props}      
\centering                          
\begin{tabular}{lll}        
\hline\hline                 
alias & ESO\,350-IG38 & (1) \\ 
coordinates & 00$^{\rm h}$36$^{\rm m}$52$^{\rm
s}$.7 $-$33\degr33$^{\prime}$17$^{\prime\prime}$ & (1) \\
systemic velocity & 6175 \kms & (1,2) \\
distance $D$ & 82 Mpc & (1) \\ 
metallicity 12+[O/H] & 7.9 & (3) \\ 
$M_B$ & $-$20.06 & (3) \\
$B-V$ & 0.58 & (3) \\ 
$V-{K}$ & 2.33 & (3) \\
$L_{\rm FUV}$ & 10$^{10.3}$ \lsun & (7) \\
$L_{\rm FIR}$ & 10$^{11.1}$ \lsun & (7) \\
$L_{H\alpha}$ & $3.0\times10^{42}$ erg/s/cm$^2$& (8) \\
$M_{\rm \sigma(H\alpha)}$ & $1.9\times10^{10}$ \msun & (4) \\
$M_\star$ & $1.6\times10^{10}$ \msun & (4) \\
\hline 
\end{tabular}
{\footnotesize \raggedright 
References: (1) NED, 
(2) \citet{Bergvall00},   
(3) \citet{BO02}, rescaled to $H_0=73$, 
(4) \citet{o01}, 
(7) \citet{o09} 
(8) \citet{Hayes07}
}
\end{table}

\begin{table*} 
\caption{Log of spectroscopic observations}\label{tab-observations}
\vspace{\baselineskip}
\begin{tabular}{@{}llllrlr}
Date &   Seeing  & Instrument & Wavelength & Resolution & Aperture & Total integra- \\
(UT) &   (\arcsec) & & coverage (\AA)   & (\kms)  &  & tion time (s) \\
\hline
1999 09 03 & 1.0 & 3.6m/CIGALE & 6693-6704 & 29 & 256x256 pixels (0\farcs 94 each)& 3180\\
2003 08 06 &  0.7 & VLT/FORS2& 7675-9420  & 63 & slit \kb\ (through knot \kb),  $PA=140$\degr\   & 6075 \\
2003 08 07 &  0.7 & VLT/FORS2 & 7675-9420  & 63 & slit \sca\  (through \kc\ and \ka), $PA=57.5$\degr\  & 5400  \\
2003 08 07 &  0.65 & VLT/FORS2& 3330-6120  & 160 & slit \sca , $PA=57.5$\degr\  & 180 \\
2003 08 07 &  0.65 & VLT/FORS2& 5120-8450  & 147 & slit \sca , $PA=57.5$\degr\  & 180 \\
2004 11 06 & 0.65 & VLT/FLAMES/ARGUS & 8180-9340 & 29 & 22x14 element IFU (0\farcs52 each), PA 90\degr\ & 14400 \\
2004 10 07& 0.7 & VLT/VIMOS/IFU & 4150-6200 & 124 & 40x40 element IFU (0\farcs66 each), PA 45\degr\ & 7100 \\
\hline
\end{tabular}
\label{obslog}  
\end{table*}

\section{Observations and reductions}
\label{sec:observ-data-reduct}

In this paper we make use of several spectroscopic observations of \haro\, all obtained at the ESO telescopes in Chile;
for a summary see Table \ref{obslog}.
In addition we show  images from the NASA/ESA Hubble Space Telescope (HST), obtained under
general observer programmes 10575 and 10902 (PI \"Ostlin).

\subsection{Long-slit spectra with FORS2}

Our long-slit spectra were taken with FORS2 (under ESO programme  071.B-0602) at moderate and low
resolution, using two slit orientations on the galaxy, see Table \ref{obslog} and Fig. \ref{nic3}. 
One, at position angle (PA) 57.5\degr\ passes
through both knot \kc\ and knot \ka\ (hereafter referred to as slit \sca); 
the other at PA 140\degr\  (hereafter referred to as slit \kb) passes       
through knot \kb\ and the galaxy's apparent kinematical centre
\citep[see Fig. \ref{nic3} and][]{Hayes07}. We obtained spectra for both position angles using
the high resolution G1028z grism and a slit width of 0.7\arcsec, and for slit \sca\ only also using the
lower resolution  G600B and G600RI grisms.
 The higher-resolution spectra cover $\sim 7700-9400$\AA\ and  include several transitions of the \hi\
Paschen series, nebular lines such as \siii 9069 and the \caii\ triplet. The lower-resolution data cover
the spectra region from 3330 \AA\ to 8450 \AA.
The relatively narrow slit width was chosen to improve spectral 
resolution ($R\sim 4800$ and 2000, for the high and low resolutions grisms respectively). 
A number of template stars (of spectral type F8 to K2), to be used in the 
cross-correlation analysis, were also observed with the same setup. The instrument
resolution was measured from unresolved sky lines.

 The low resolution spectra were obtained and reduced in a standard manner 
with wavelength calibration provided by arc lamp exposures.

The G1028z observations were designed to optimise  subtraction of the strong 
sky emission from
atmospheric OH and O$_2$. 
For both slit positions, we took a number of identical
exposures with the galaxy placed at different positions along the slit.
The  nodding step between two consecutive exposures were set to 
always 
have different length   and being larger
than the apparent size of the galaxy.

First, a constant background value was subtracted from each exposure, 
with the level determined between the OH lines at 8510 \AA.  Then
adjacent pairs of exposures were subtracted from each other, each time
scaling the subtracted frame so that residuals from the OH and O$_2$
lines were minimised.  
Wavelength
calibration was carried out in two dimensions using  OH sky
lines  in the non subtracted frames, and wavelengths from \citet{o96}.
   To remove pixel-to-pixel variations in the response of the CCD, we
divided each pair-subtracted frame by a dome flatfield. 
We found that the OH and O$_2$ lines
varied somewhat differently from each other from exposure to exposure.
They are, however, easy to separate.  In the region of interest, O$_2$
emission dominates the background between 8600 and 8720 \AA, so we
were able to perform scaling and subtraction for the two cases
separately, and then spliced the spectra together again after
flat-fielding.  The spectra were registered and median combined
to produce the final frames for use in the subsequent analysis.
 We extracted the spectra three detector rows at a
time (0\farcs 75) and flux-calibrated by comparison with standard
stars LTT 7379 and LTT 7987.

\subsection{Integral field spectra with FLAMES/ARGUS}

Integral field unit (IFU) spectroscopy was taken using the FLAMES integral field
unit ARGUS, connected to the spectrograph GIRAFFE \citep{Pasquini02} under observing programme 074.B-0771.
We used the $0\farcs52$ pix$^{-1}$ scale with grating setup LR8 giving a spectral coverage  
from $8200$  to $9380$\AA\ with spectral resolution $R\approx 10\,400$.
The $22\times14$ lenslets result in a field-of-view of
$11\farcs4 \times 7\farcs3$ placed with the long side in east-west
direction  (see Fig. \ref{nic3}).  
Three spectral template stars (of spectral type K0II, K0III, G8Iab) were 
observed with the same setup.

 The data were
reduced in the same manner as in \citet{m07} using the
software and methods of \citet{2002A&A...385.1095P} for bias
subtraction, flat-fielding, wavelength calibration and optimised
extraction of each spectrum. Sky emission was subtracted using
simultaneously observed spectra from the separate 15 sky fibres  \citep[see
  \eg\ ][]{2003Msngr.113...15K}. As a consistency check, we integrated
the spectra over the whole galaxy and derived the systemic velocity to
be 6179 \kms, in good agreement with previous results (cf.~Table
\ref{table:props}).

\subsection{Integral field spectra with VIMOS}

\haro\ was also observed with the VIMOS IFU using the HR-blue grism in 
program 74.B-0802 (PI Bergvall). These observations, originally designed 
for a different scientific purpose, were obtained at a pointing well 
outside the main body (15\arcsec to the NW, see Fig.  \ref{nic3}), 
but contain velocity information on the ionised
gas that was deemed useful for the current investigation, and therefore 
these data were included in our analysis.

The VIMOS/IFU with    HR-blue grism\footnote{Note that this is the old HR-blue grism. The new HR-blue grism 
has lower spectral resolution and a different wavelength range. } and scale $0\farcs67$ per fibre
delivers a field of view of $27\arcsec\times27\arcsec$ over the
wavelength range 4150--6200 \AA. Of interest for the present work is
that we  detect \oiii $\lambda\lambda4959, 5007$ and often also
\hb\ over most of the field of view. The spectral resolution is
$R\approx 2550$, verified by the width of sky lines in the spectrum.
The IFU spectra results from a single 7200s integration. 
 In order to remove cosmic rays in this single deep exposure, we first normalised each 
spectrum to the continuum near \oiii\ and took the median of $5\times5$ pixels,   constructing
a binned spectral image with  3\farcs3 resolution. Since VIMOS/IFU lack
dedicated sky fibers, the sky level was estimated from a dedicated blank sky exposure
of 1800s next to the target.  
Some regions of the IFU showed a significantly higher residual background level and were 
not included in the further analysis.

\subsection{Fabry-Perot interferometry}

Fabry-Perot interferometry  was collected  with  CIGALE \citep{Amram91} mounted 
on the ESO 3.6-m telescope, under program 63.N-0736, using an etalon with free spectral 
range (FSR) of $  390$ \kms . The spatial sampling was 0.94\arcsec/pixel. The etalon was scanned in 64 channels (each 6.1 \kms\ wide) for a total integration 
time of 3200s. The data was reduced with the {\sc Adhoc} \citep{Boulesteix93} 
software, and a Gaussian smoothing of 1\farcs 5 was applied. The effective 
spectral resolution in the reduced data cube as measured by the width of the instrument 
profile is 29 km/s FWHM.

\section{Stellar kinematics from the Calcium triplet}
\label{sec:stellar-kin}

Observing the stellar velocity field in metal-poor starbursting galaxies
is very challenging due to  intrinsically weak stellar absorption lines
and contamination from ionised gas emission.  In galaxies like \haro\ with very strong emission
line spectrum, the Balmer  lines become completely emission dominated.
The Calcium triplet, which originates from photospheric 
absorption in red giants and red super giant stars, is one of very few practically accessible probes.
However the Calcium triplet lines are partly blended with Paschen lines which needs to be dealt with
before stellar velocities can be derived.

\subsection{Removing Paschen emission lines}

We subtracted the lines of the \hi\ Paschen series from our FORS2
spectra according to the scheme presented in \citet{c08}.  As
a template for the higher order Paschen lines, we constructed a model
profile by adding the profiles of the three strongest \hi\ lines in
our spectra   that are  not blended with the Calcium triplet lines (Pa\,10, 11 and 12).
Where the
signal to noise ratio was low, we  used \siii\ $\lambda$9069 as
a template. 

We adjusted the width, flux and velocity of the model lines in order
to give the best subtraction around those Paschen lines which do not
blend with \caii\ absorption. Finally we masked the few other
emission lines seen in the spectra, \Oi \ $\lambda8446$, [Cl~{\sc ii}]
$\lambda8579$ and \feii\ $\lambda8617$.

For the ARGUS spectra, we followed the same procedure to subtract
Paschen emission lines, using Pa\,10 as the sole profile template due to
the limited wavelength coverage.

   \begin{figure*} 
   \centering
    \includegraphics[angle=0]{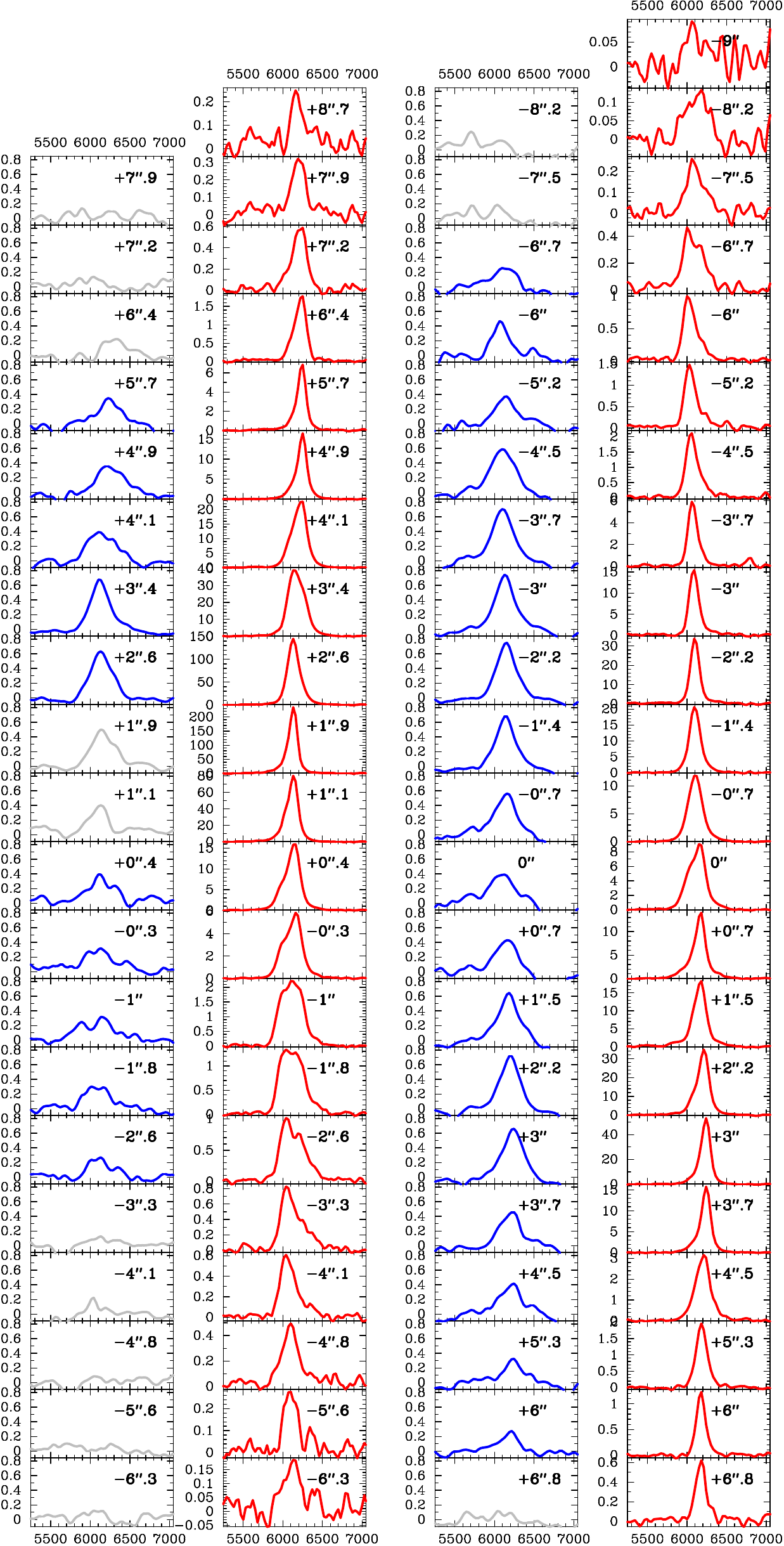}
  \caption{Variation along the slit of the cross-correlation function (CCF, first and third columns) from {\sc fxcor} 
   and the  \siii\  $\lambda 9069$ line profile (second and fourth columns). The leftmost two columns show the 
   results for  slit \kb\  and the rightmost columns
   the results for  slit  \sca . Where the CCF peak value is smaller than 0.25 or the Paschen-subtraction
   is deemed unreliable, the CCF  is shown in grey. The emission-line
   profiles have been continuum-subtracted.
   Each sub-panel is labelled with the position of the extraction window, where the origin
   is set to the position where the two slits cross (see Fig. \ref{nic3}). Knot
   \kb\ is at position $+$1\farcs9 along slit \kb , and knots
  \kc\ and \ka\ are  at $-$2\farcs2 and $+3$\arcsec\  
   along slit \sca , respectively.   }
         \label{f-ccfs}
   \end{figure*}

\subsection{Stellar kinematics from FORS2 slit spectra}
\label{s_sk}
\begin{figure*} 
   \centering
 \includegraphics[angle=-90,width=1\textwidth]{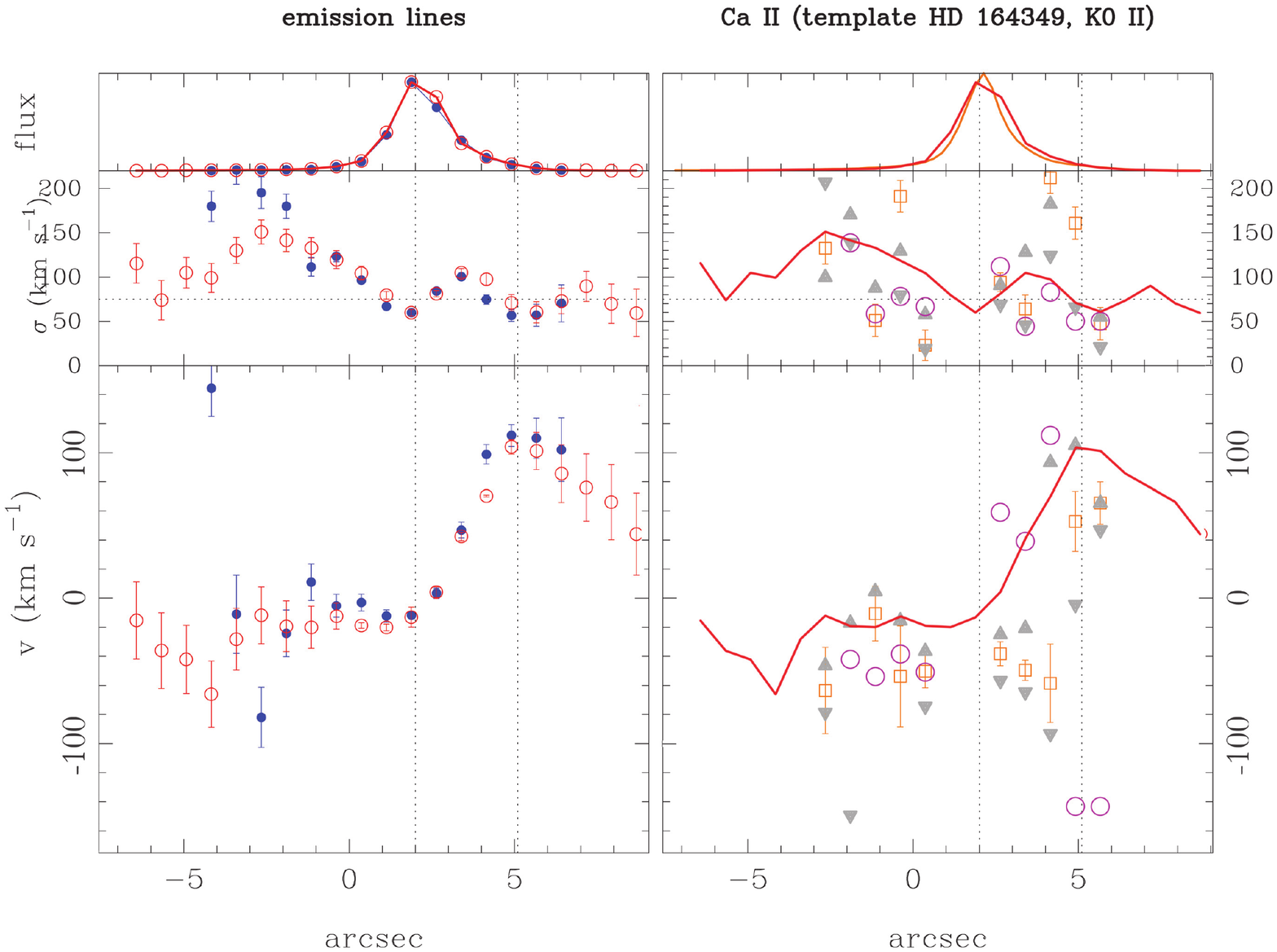}
   \caption{Kinematical results along FORS2 slit \kb\ showing (from top to bottom) 
   the flux level, the line of sight velocity dispersion, and the heliocentric radial velocity (relative to the systemic velocity of 6175 \kms ).
   Zero on the abscissa is defined as the point where our two slits cross.  
   The {\em left} panels show results for the  \siii\ $\lambda$9069 (red open circles), 
   and \hi\ Paschen emission lines (weighted means of all measured lines; blue filled
   circles).
   The flux scale show the normalised strengths of the emission lines.
    The dotted vertical lines at 2\arcsec\ and 5\arcsec\ marks
   the position along the slit of knot \kb\ and where the slit crosses the loop of structures 
   referred to as the 'ear' (see also Figs. \ref{nic3} and \ref{cigale}), respectively.
   The dotted horisontal line in the $\sigma$-plot is arbitrarily put at  75 \kms .
   The {\em right} panel shows at the top the spatial variation of the normalised continuum flux near the Calcium triplet 
   (orange), in arbitrary units. The red solid line repeats the \siii\ flux distribution 
   from the left panel. The middle and bottom panels show the derived stellar velocity dispersion and radial
   velocity, respectively. Again, the solid red line repeats the \siii\ results from the left panels. The results from  
   cross-correlation is shown as orange squares and the pPXF results 
   as magenta open circles. The grey triangles represent the
extreme values derived (as described in
sect. \ref{s_sk}) from the cross-correlation method when systematics in the Paschen emission line subtraction is taken into account.
   At the
   distance of \haro,  10\arcsec\ corresponds to 4 kpc.
   }
      \label{f-rc-1}
   \end{figure*}

\begin{figure*}   
   \centering
   \includegraphics[angle=0,width=.92\textwidth]{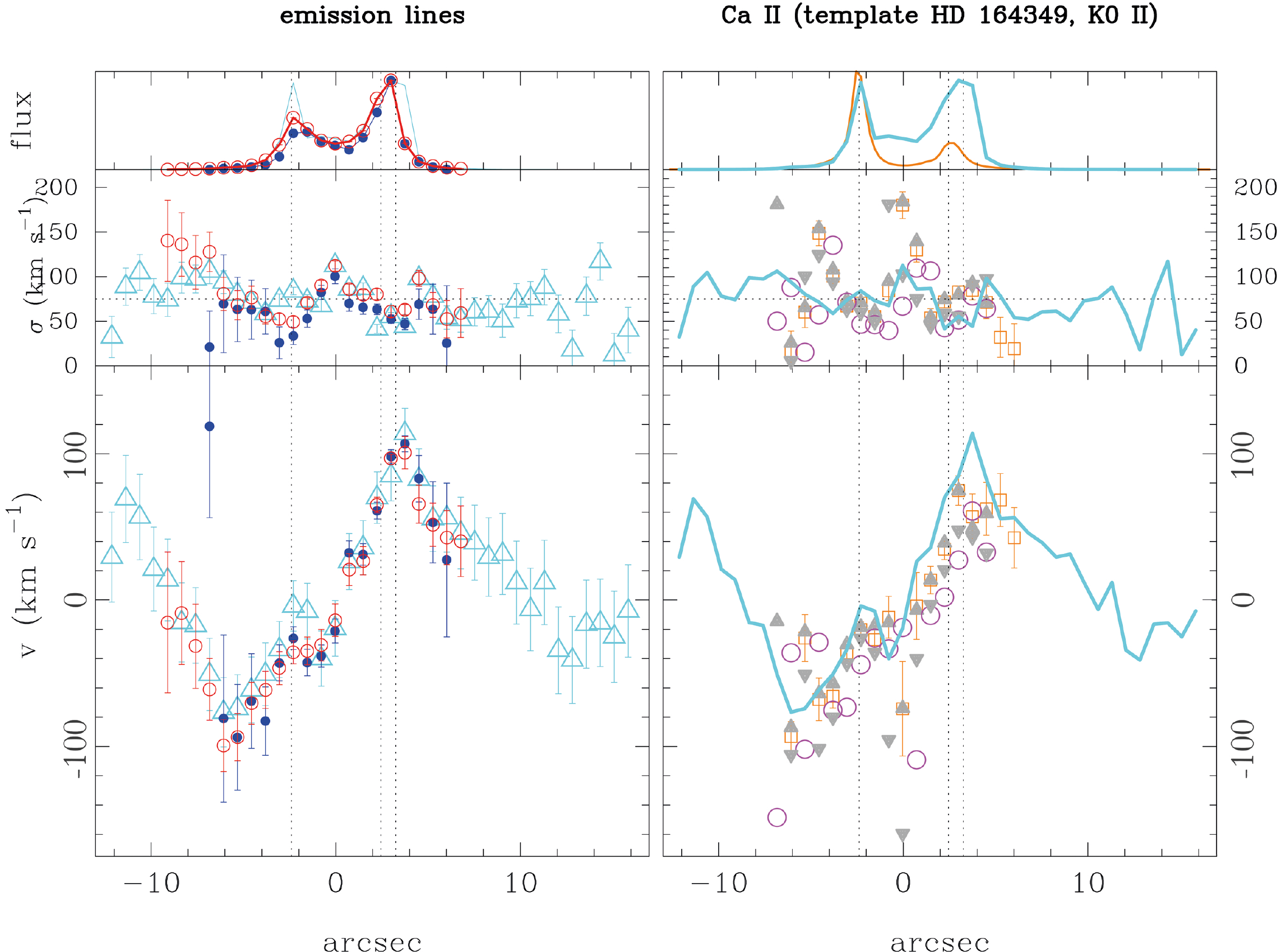}
      \caption{Same as Figure \ref{f-rc-1}, but for slit \sca , and with the addition of \ha\  measurements (cyan triangles). 
      In the right panels, the cyan full drawn line repeats the \ha\ measurements from
      the left panels, to allow for comparison of the stellar and ionised gas kinematics. Dotted vertical lines mark the positions of
      knots \kc\ (left dotted line) and  \ka\ (two rightmost dotted lines), see Fig. \ref{nic3}.}
         \label{f-rc-2}
   \end{figure*}

 We derived the stellar kinematics by cross-correlation of the cleaned spectra 
using  FXCOR \citep[based on][]{TD} and the same set of template stars as in \citet{c08}.
We  followed the method of \citet{o07} for deriving the resulting errors in the velocity
dispersion.
Figure \ref{f-ccfs} shows how the cross correlation function 
(CCF) peak and the \siii\ profiles vary along each slit. 
In the right panels of  figures \ref{f-rc-1} and \ref{f-rc-2} we show the 
resulting stellar velocities  (relative to the systemic velocity 6175 \kms ) and velocity dispersions (with orange squares plus
error bars).
Independently, we derived the velocities and dispersions using the penalised
pixel-fitting (pPXF) method of \citet{CE04} (shown as magenta circles in Fig. 
\ref{f-rc-1} and \ref{f-rc-2}).

The derived velocities and velocity dispersions are subject to
systematic errors from the Paschen line subtraction, in
particular where the Paschen emission is strong and the Calcium
triplet weak. In Fig \ref{f-ccfs} those CCFs for which the peak amplitude was below 0.25, or  the
Paschen-subtraction was not deemed reliable, are shown in grey, and
these points
were discarded from the subsequent  analysis. To quantify the effects of 
 over- and
under-subtraction of the Paschen lines, we additionally investigated two extreme cases: \\

\noindent (i) We increased the Paschen-line intensities as much as
possible without inducing artificial absorption features , all the while adjusting the
velocity dispersion and velocity of the nebular line model to allow 
the maximum intensity to be subtracted. 

\noindent  (ii) In the other case we instead kept the
subtracted intensity as low as possible without leaving any residual
emission features at the unencumbered Paschen lines, again letting the
velocity and velocity dispersion take any values to assure this.

The spectra resulting from the  two alternative  subtractions were then cross-correlated as before.
The results bracket the possible range  of velocities allowed by the data, and give a good estimate of the maximum systematic errors. 
In figures \ref{f-rc-1} and \ref{f-rc-2} we show, with upwards and downwards 
triangles, the  velocities and dispersions resulting from  these extreme
Paschen subtractions. 
The comparison with the pPXF results give further
information on the systematics. Sometimes the differences 
are larger than those represented by the error bars, and other times smaller. 
In the latter case  the obtained result should be regarded as quite robust.
The stellar
velocities are found to be insensitive to the template star used in
the cross-correlation, so we treat the data for the bright KO  giant
HD\,164349 to be representative.

For slit \sca\ we see that the stellar velocities and dispersions resulting from the
two different methods (FXCOR and PPXF) and extreme Paschen-subtractions 
agree very well, and hence the overall velocity and dispersion profile must be
regarded as very reliable. For slit \kb\ the agreement is worse, partly due to the 
emission lines being stronger along this slit.

\subsection{Stellar kinematics from FLAMES/ARGUS}

After Paschen-subtraction we followed the procedure described in \citet{m07}. 
However, poor signal-to-noise and strong Pa emission lines in many of the 
spaxels meant that we could not derive a detailed spatial map of the stellar 
velocities. After binning 2$\times$2 we could determine the stellar velocity in 
the region around \kc\ where the Paschen emission is relatively weak, finding 
good agreement with the FORS2 result. The only region
where we get reliable velocity data and that was not covered by by FORS2 is 
2\arcsec\ South-East of knot \kc. Here the stellar velocities are the same as
those for the gas (derived from \siii ). We will  not discuss the stellar kinematics 
from ARGUS further.

\begin{figure*}  
  \resizebox{\hsize}{!}{\includegraphics{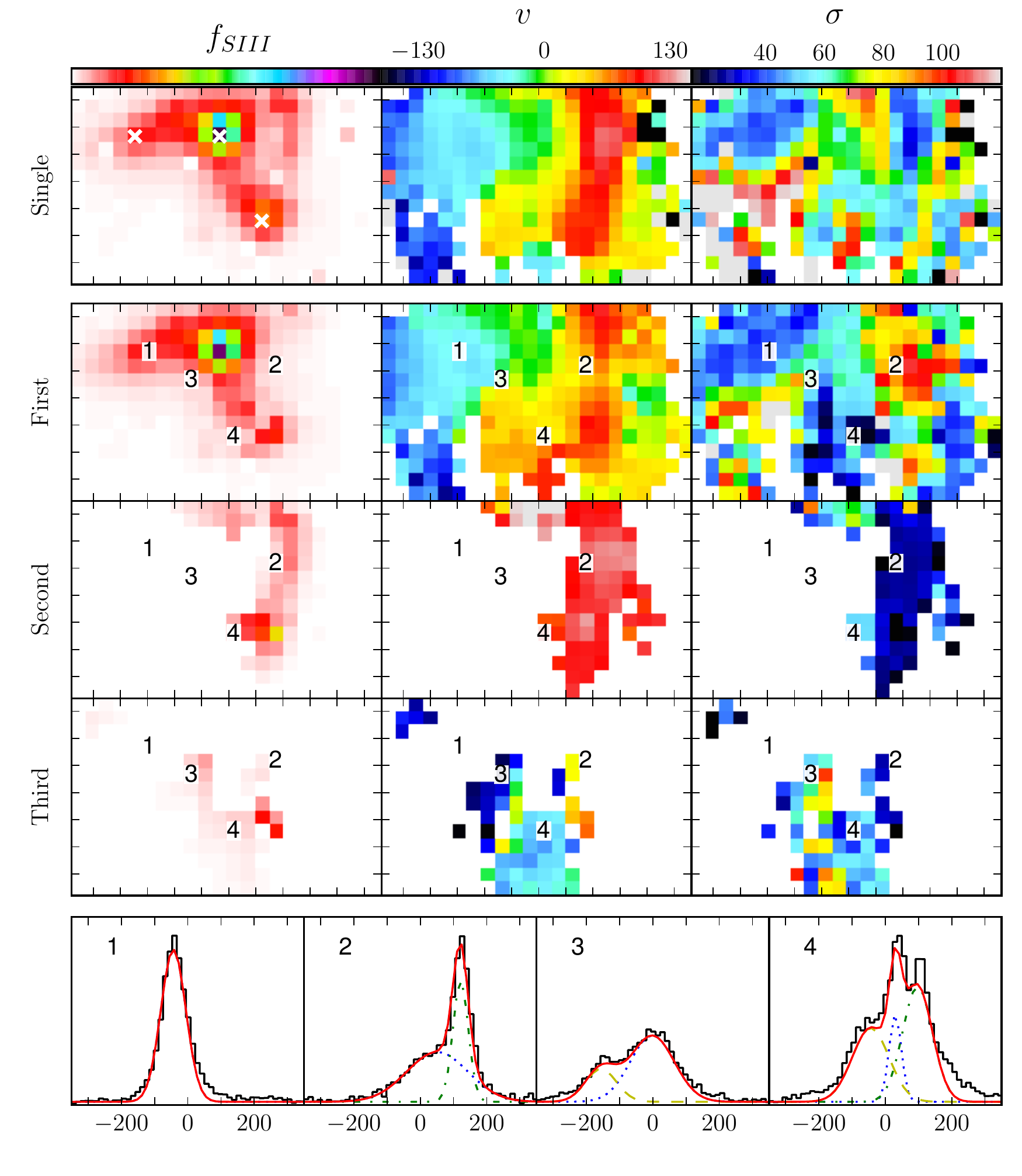}}
  \caption{Maps of the field-of-view of FLAMES/ARGUS showing the
    derived properties of the \siii\ $\lambda9069$ line. The tick marks
    have a spacing of 1\arcsec. The {\em top row} shows the
    results from a single component fit, from left to right: intensity
    in arbitrary logarithmic units (with the location of the 3 knots indicated with  crosses),
     velocity and     velocity dispersion. 
     Since we find that over large areas,  a single velocity component is
    insufficient to reproduce the line profile,    we present an attempt to decompose the line
    by fitting multiple gaussians. The number of components was chosen
    conservatively and  additional components  were only  included in  when
    deemed necessary to produce a decent fit. The results are presented in the
    {\em three middle rows}.       The \emph{bottom row} shows examples of
    measured \siii\ line profiles (black solid line), and our
    multi-component fits to them (first: blue dotted, second: green
    dash-dotted, third: yellow dashed). The red solid line is the sum
    of the components. The x-axis displays velocity, relative to the
    systemic velocity, in \kms. The positions that correspond to the
    spectra are shown by numbers over-plotted on the panels above
    them.}
  \label{argus-siii}
\end{figure*}

%
%
\section{Ionised gas kinematics}
\label{sec:gas-kin}

Velocities for the ionised gas are calculated using rest wavelengths from 
   \citet{AtomicLineList} and (for \siii\ only) \citet{HM81}. 
   
\subsection{Gas kinematics along the FORS2 slits}
The analysis of the FORS2 ionized gas velocities and their errors follow  \citep{c08}.
In general we use \siii\ $\lambda$9069, the strongest line in the G1028z
spectra. Along slit position \sca\ we also observed   \ha\ with grism G600RI, which 
allow us to probe the velocities to somewhat
larger radii than for \siii , albeit with lower spectral resolution.
The   \siii\  emission line profiles are shown in Fig \ref{f-ccfs} and 
it is obvious that they are non-Gaussian at many places.  
Since the ARGUS data has higher spectral resolution we restrict our analysis of 
the detailed line shape to this data, see next subsection. 
In Fig. \ref{f-rc-1} and \ref{f-rc-2} we show the velocities and
velocity dispersions for stars and gas derived from the assumption of a 
single Gaussian component.  We note that the regions coinciding with
knots \ka, \kb\ and \kc\ have smaller velocity dispersion than the average of their surroundings.

   \begin{figure*}
   \centering
  \includegraphics[angle=0,width=0.751\textwidth]{haro11cigale_new.pdf}
   \caption{Results from CIGALE Fabry-Perot \ha\ observations of \haro. \emph{Upper left:}
      Line-of-sight velocity (relative to the systemic velocity)  with the scale (in \kms )
     given by the colourbar beneath. The
     spatial scale is indicated in the corner and a
     single \ha\ flux contour (at $\sim5\cdot10^{-15}$\flux arcsec$^{-2}$) is over-plotted for reference. 
     The solid black line shows the line along which the rotation curve (in the lowe left panel) was extracted 
     and the dashed lines indicate the $\pm 40\deg$ wedge of points included.  
        \emph{Upper right:}
     Velocity dispersion map. The same contour as in the first panel
     is over-plotted. The white circles  mark the positions of knots \ka, \kb
     and \kc , and  diamond marks the ``overlap
     region''. \emph{Lower left:} 
     Rotation curve, derived  by deprojecting the
     apparent position of each velocity point to its radius in the
     inclined plane, using the position angle and wedge indicated in the upper left panel.
     The error bars correspond to the spread of the
     velocity measurements within each one arcsecond wide radial
     bin. \emph{Lower right:} Example of \ha\ profiles from the locations
     of the three knots and the ``overlap region''.}
         \label{cigale}
   \end{figure*}

\subsection{ARGUS}
We used \siii\ $\lambda$9069, the strongest emission line in our spectra from ARGUS,
to derive spatially resolved kinematics of the
ionised interstellar medium. The top row of Fig.~\ref{argus-siii}
displays maps of amplitude, velocity and velocity dispersion resulting
from a fit to the line using a single velocity component. Because the
line shape at this spectral resolution is non-Gaussian over large regions of the field-of-view, we
allowed for two additional degrees of freedom in the form of the
parameters $h_3$ and $h_4$  from the Gauss-Hermite polynomials. They
represent skewness and   kurtosis (spikiness/boxiness) respectively and generally
allow the fitted velocity to better correspond to the peak of the line
profile.

The \siii\ line shows a complex kinematic structure and several individual
line profiles indicate the presence of multiple velocity components. We
next decomposed the \siii\ line profiles into
several components by fitting multiple Gaussians in the regions where
it is clear that the profile cannot be described by a single component. We use
the smallest numbers of components deemed necessary from the 
shape of the line profile and the quality of the data. This means that 
for some pixels we have one component, for some two and for others
three. Even three components is sometimes insufficient (e.g. lower right
panel labelled '4' in Fig.~\ref{argus-siii}) but we refrain from fitting more
that 3 profile due to the ambiguity in the interpretation.

The line  decomposition in some places reveal up to three dynamical components 
-- how can we {\em spatially} make sense of these components?
There is a principle problem of multi-component fits in that one needs
to arrive at a spatially consistent identification and labelling. The question
whether a component at one place in the galaxy is related to a particular
 component in a nearby region is non-trivial as each
component may vary spatially both in velocity, width and
amplitude. We use spatial continuity over several pixels as a criterion for manually 
deciding on how to sort the components in an iterative fashion. This means that 
we tried to avoid discrete jumps in velocity and line width for a given
component. 
Rows 2-4 in Fig.~\ref{argus-siii} display maps of the resulting kinematical 
components: the {\em first}  component is wide spread over the field. 
The {\em second} component forms a kinematically cold 
structure that is well defined in space, intensity,  velocity and line width and which
we associate with the "ear".
The {\em third} component is somewhat more elusive. The first component 
represents everything that does not belong to the second or third component
or when the line shape did not obviously require more than one component.

In addition to the sorting problem, there is another uncertaintiy that affect the 
velocities and dispersions shown in Fig.~\ref{argus-siii}:   
Multiple components can only be clearly identified if they have different amplitude or 
are separated in velocity by an amount comparable to the FWHM
of the individual lines. This means that even where a single component may seem sufficient, 
there might in reality be several distinct regions projected along the line-of-sight. 
If the velocity separation and relative intensity of different components change, at some 
positions they may blend and two relatively narrow slightly separated components may be 
mistaken for a single broad one, leading to velocity and $\sigma$ jumps.

\subsection{CIGALE}
The \ha\ velocity field from CIGALE was analysed by fitting a Gauss-Hermite 
polynomial to the line profiles in each pixel.   The resulting velocity field  ($h_1$) and velocity 
dispersion ($h_2$) are shown in Figure \ref{cigale}. There is a good agreement with
the \siii\ velocity field (Fig. \ref{argus-siii}). In order to allow comparison with the
results of \citet{o99,o01} we have also extracted a rotation curve along the
major axis defined by the new photometry presented in \citet{Micheva2010}, where
the shape of the  $K=23$ mag/arcsec$^2$   isophote was fitted with a position angle  
$PA=103\degr$ and an
     inclination  $i=47\degr$ (see Fig. \ref{cigale}). 
The line of sight \ha\ line widths are $\sigma\approx60$ \kms\ at \ka\ and \kc,
and $\sim80$ \kms\ at \kb. The broadest lines are seen in the 'overlap' region
 (i.e.  approximately the point where the two FORS2 slits cross and where the ARGUS 
data reveals at least 3 velocity components in the \siii -line), 
and also 4\arcsec\ south of this region and in addition 2-3\arcsec\ north and east of \kc .

\subsection{VIMOS}
In the  VIMOS data, the \hb , \oiii\ $\lambda$4959, and
$\lambda5007$ lines are detected in all binned pixels. We simultaneously
fitted the \oiii\ lines with a single redshifted Gaussian each, connected
by the intensity ratio 1:3, a common width and a separation of
$\Delta\lambda=(5007-4959)(1+v_r/c)$ \AA\ where $v_r$ is the fitted
radial velocity. We independently fitted the \hb\ line with a single gaussian.

\begin{figure}
   \centering
     \includegraphics[clip=true, trim=5 0 -15 0,angle=0,width=.5\textwidth]{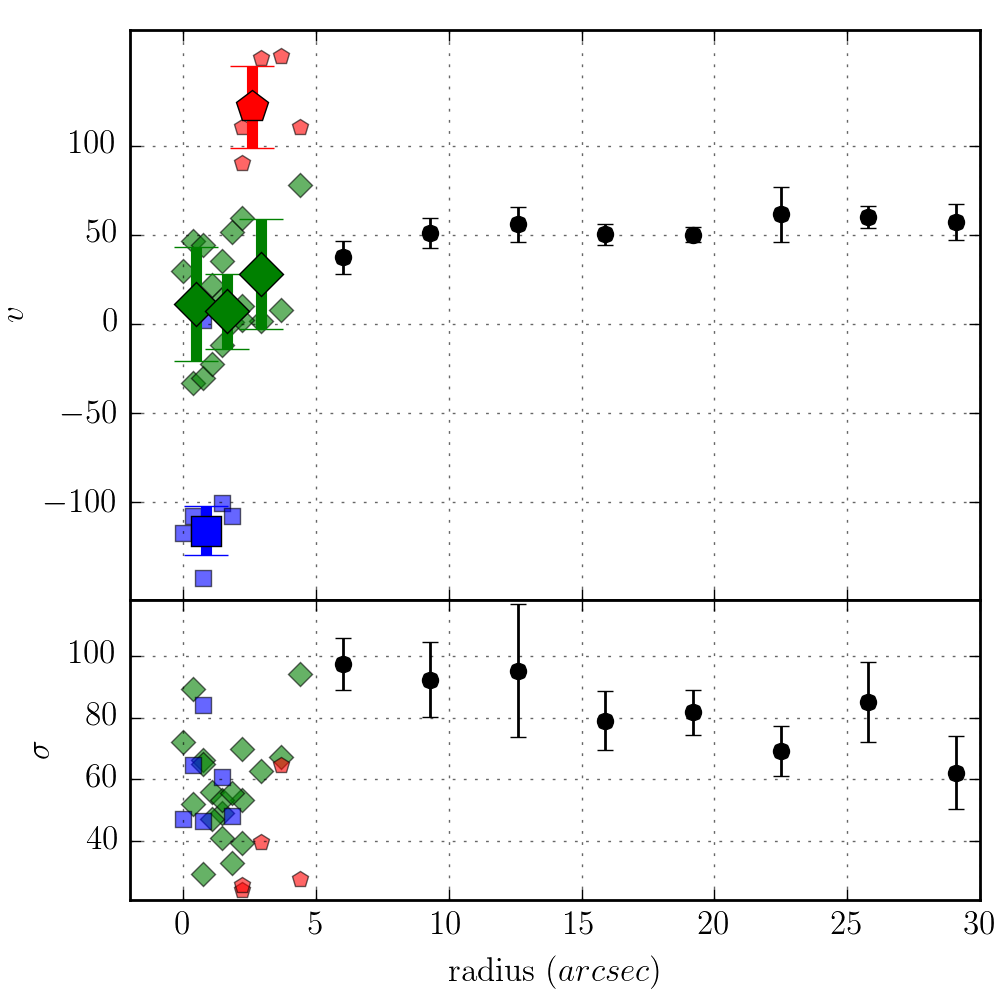}
 \caption{Line-of-sight velocity of gaseous emission lines along PA
     -45\degr . Black dots are the
     \oiii\ velocities and line of sight $\sigma$ from our VIMOS data. The errorbars for all
     points represent the spread (1$\sigma$) of the individual
     measurements within the radial bin, not the measurement error.
     Green, red and blue dots correspond to the first,
     second and third velocity component from the ARGUS analysis of the \siii\ line
     (cf.~Fig.~\ref{argus-siii}), as derived by a cut through the
     velocity field in the north-west direction. Small symbols represent individual spaniels and 
     large symbols radial averages with error bars show the  spread. The scale is 0.4 kpc/arcsec.}
         \label{vimos}
   \end{figure}

The field-of-view of the VIMOS IFU was placed at a 45\degr\ angle
north-west of knot \kb .
 Since we are interested in how the gas
velocity and line width changes with distance from the starburst, we 
collapsed the
velocity field along the direction of slit B, resulting in a
single set of velocity versus radius data, shown as black dots in
Fig.~\ref{vimos}. Scattered light could potentially be an issue with
VIMOS IFU data \citep{Monreal06}, however, our pointing is far from the
bright centre which is well outside the field of view of VIMOS/IFU.

The \hb\ data is quite noisy but on average the \oiii/\hb\ ratio is $\sim 5$,
similar to what is seen in the centre (Bergvall \& \"Ostlin 2002).
We find that the \hb\ velocities to match those of the \oiii\ lines within the uncertainties. 
Therefore we only further investigate the latter.
The \oiii\  lines show no apparent structure, such as asymmetries or
double components. They are however, significantly broader than the 
instrument profile, which we measured to be $FWHM_{\rm instr}=124$\kms
(or equivalently $\sigma_{\rm instr}=53$\kms) .
 In the high signal to noise bins, the raw width
as determined from a Gaussian is $\sigma_{raw}=100 \pm 10$ \kms , 
implying an instrument corrected line width of $\sigma \approx 85 \pm 10$ \kms , 
i.e. very similar to the average value in the centre of 81 \kms \citep{o01}, see also Fig. \ref{cigale}.
There is a weak decline in $\sigma$
 with increasing radius.

%
%
%
\section{Results and analysis}
\label{sec:results}

Figures \ref{f-rc-1} and \ref{f-rc-2} show how the velocity, $\sigma$ 
and flux vary along the FORS2 slits, for selected 
emission lines (left panels) and for \caii\ (right panels). 
The results for slit \sca\  (Fig. \ref{f-rc-2}) show a clear correspondence between 
the velocities and dispersions for gas and stars. 
The velocity curves are not identical, but overall, stars and gas have very
similar kinematical properties. Slit \kb\ adheres to this trend even if the poorer
data makes it less obvious. 

The ARGUS ionised gas velocity field on closer inspection showed evidence
for multiple kinematical components, as discussed below.
Also the stellar velocities may be multi component in nature  as hinted by the shape of some of the CCFs in Fig. \ref{f-ccfs}, but the current
data does not allow us to reliably investigate this further.

The similarity of the stellar and gaseous velocities suggest that have a similar
physical origin, namely virial motions. Whereas gas is known to be affected by
feedback, this apparently has not affected the line core, perhaps because 
gas that is affected by feedback tends to become diluted and have low intensity.
The motions of stars are, on the other hand, not affected by feedback.

\subsection{Analysis of the \siii\ velocity field}

The high spectral resolution of the ARGUS IFU-spectra allows us to
have a closer look at the shape of a gaseous emission line. If one
measures a line shape that is clearly the sum of multiple components
with different velocities, it is safe to assume that they originate
from physically distinct regions since it is hard to imagine a single
region with two or more distinct bulk motions. This means that we have
a handle on the third dimension of \haro, perpendicular to the
 plane of the sky. 

 Fig.\ref{argus-siii} shows the result and some
examples of fitted line profiles. Several results emerge from this:

(1) There is a main velocity component in the sense that it can be
seen across the galaxy. There seems to be no indication of a
discontinuity as one would expect in a system that has not yet merged.
The decomposed line has its broadest ($\sigma\approx90$\kms) 
component $\sim1\arcsec$ south-west of knot \kb\ (see spectrum '2'
in Fig.\ref{argus-siii}). This is also the apparently most dusty region {\bf (see Fig. \ref{ant}).}
It is possible that the broad component giving rise to the high line
width is in fact due to unresolved narrower components.

(2) The second velocity component show a near spatial coincidence with 
the 'ear' (Fig. \ref{nic3}). It forms a high velocity component that is narrow
in space and velocity ($\sigma\sim30$ \kms ). It is likely a likely tidal arm, 
that was flung out  in the merger, and which may be associated with the 'ear'.

In a less detailed investigation, the high velocities at the location of
the ``ear'' could have been misinterpreted as due to rotation. Structures like
this one could easily be hiding in low spectral resolution studies of 
BCGs or merging galaxies, and highlight a potential pitfall 
in interpreting the kinematics of complex galaxies through low resolution
spectra.   

The kinematical information from VIMOS (see Fig. \ref{vimos}) support this interpretation. 
The velocities measured further to the North-West (Fig.~\ref{nic3})  do not coincide with 
the high  velocity of the second  component but instead 
joins smoothly with the first component.   Hence, the ``ear'' is a spatially and 
kinematically distinct region.
Figure \ref{nic3} shows that
the \ha\ emission in the ear is dominated by a chain of compact young star 
clusters, \citet{Adamo_haro11}.   Similar structures as the ear are seen in the Antennae
\citep[see][and also Sect. \ref{antcomp} in this paper]{w10}.

(3) There is a third lower velocity component, mostly around knot A and the "overlap" region
(position 3 in Fig.\ref{argus-siii}).

The \ha\ velocities probed by Cigale (Fig. \ref{cigale}) in general agree well with 
those from \siii\ although the spatial resolution is worse. The \ha\ line 
is markedly non Gaussian over most of the galaxy, and is broadest in the overlap
region.

\subsection{Stellar kinematics and virial masses of knots \ka ,\kb\ and \kc}
\label{sec:stars}

The stellar velocities along slit \sca\  shows
a gradient of about 50 \kms\ kpc\mone \  (Fig. \ref{f-rc-2}). 
	The stellar velocity dispersion  ranges between  60 and 150 \kms\, and 
largely follows that of the gas, peaking halfway between
knots \ka\  and \kc\  (close to 0\arcsec on the abscissa, i.e. the overlap region). 
At this point the ionised gas spectrum from all instruments (CIGALE, ARGUS, 
FORS2) is markedly multi-component. Also the CCF structure is suggestive 
of multiple stellar components, and this is the most likely explanation for the very high
stellar velocity dispersion found.

Both the stellar continuum and the absorption line equivalent width of the \caii\
triplet are strongest at the position of knot \kc\ where the velocity dispersion of 
 gas and stars show a local minimum of $\sigma \approx 50\pm10$ \kms. 
	Knot \kc\ is very compact and 
unresolved if not viewed with the HST or adaptive optics. Using the
{\tt baolab} package \citep{Larsen99} and the HST/ACS/HRC/F814W image, we
find the light distribution to be best fitted with a  King profile 
with effective radius of $r_{\rm eff}=40\pm10$ pc. 
This indicates a virial mass for knor \kc\ of:

\begin{equation}
{\mathcal M}_{\rm vir}=2.4\times10^8 \left(\frac{\sigma}{\rm 50 \kms}\right)^2 \left(\frac{r_{\rm eff}}{\rm 40 \ pc}\right) \msun.
\end{equation}

\citet{Adamo_haro11} used SED fitting to estimate an order of magnitude lower 
{\em stellar} mass  which however was based on the assumption that \kc\ was a point 
source and therefore grossly underestimated the total flux. 

 In HST images, knot \ka\ is resolved into a group of bright star clusters.
The size of the group is $r_{\rm eff} \approx 200$pc and the velocity dispersion 
 $\sigma \approx 50$ \kms . Hence the dynamical mass of the 
 {\em complex} making up knot \ka\ is 
$ {\mathcal M}_{\rm vir}\sim 10^9$\msun .

For knot  \kb\ we measure $\sigma \approx 80$\kms\ although  it is more 
uncertain than for \ka\ and \kc\  due to the very bright emission lines. 
With a size of $r_{\rm eff} \approx 140$ pc the mass is of the order $\sim2\times 10^9$\msun .

The velocity dispersion (both for gas and stars)  have local minima at the positions
of knots and bright clusters. This is understandable since in such cases
the light is dominated by the knot in question and the measured line 
width is strongly biased by the  local conditions in the knot, whereas
in a random position, the line profile has (often unresolved) contributions
from many velocity components along the line of sight. It does not
imply that the knots are necessarily dynamically cooler.

Comparing the stellar and nebular gas velocity field it is striking how similar they are on
average. While we do not have the same possibility to resolve the stellar velocity field into multiple
components, the similarities indicate that \caii\ and nebular lines trace the dynamics of
the system in a similar way. Stars are not subject to outflows and hence, while there may
well be outflow components present in the nebular spectrum at many points, they do not
dominate the local fluxes. 

   \begin{figure}
   \centering
    \includegraphics[angle=0,width=.4\textwidth]{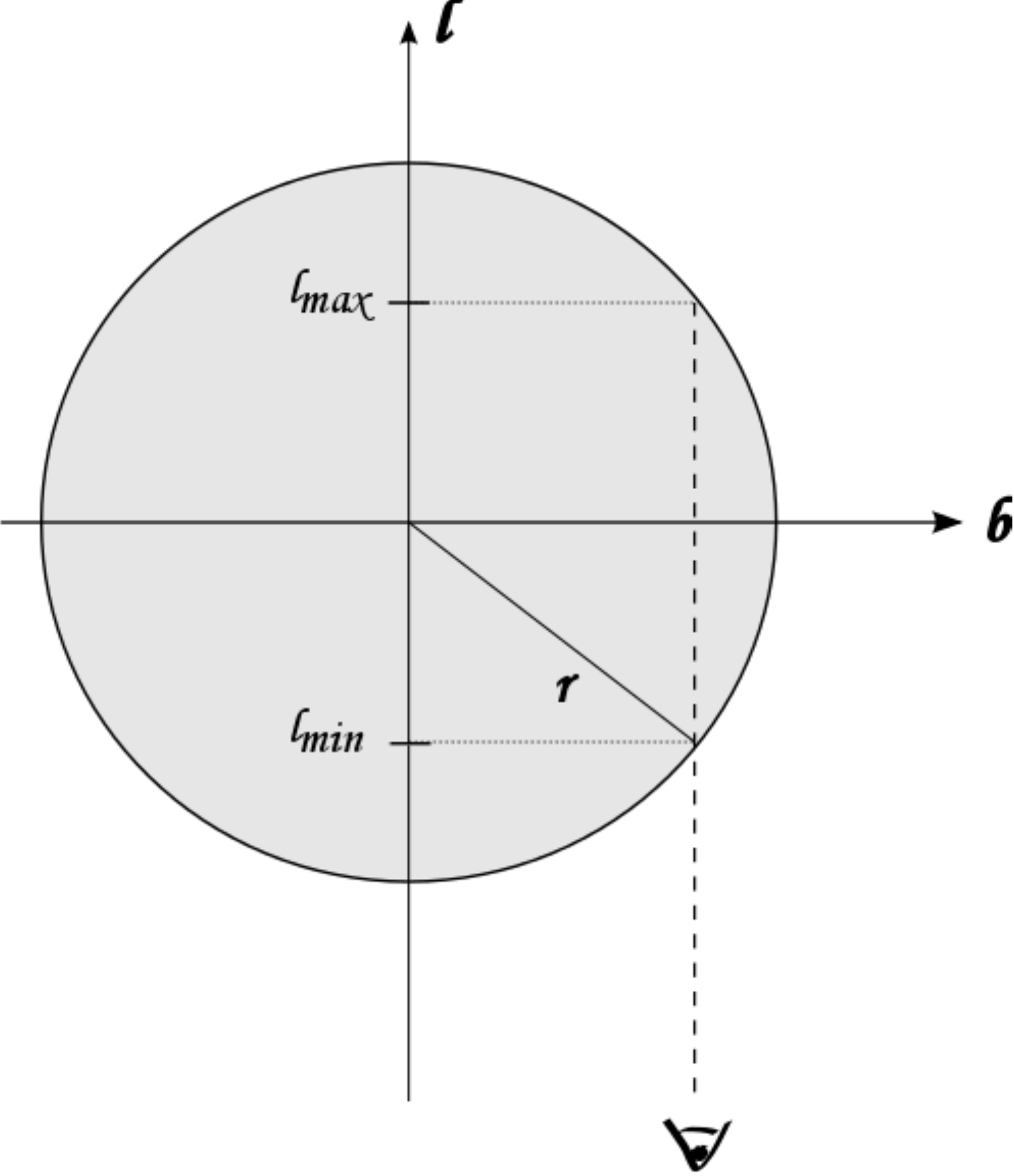}
   \caption{Schematic illustration of relevance for the discussion in \ref{sec:outflow}.
   An observer sees the galaxy at an impact parameter $b$ along the line of sight (dashed line).
   The line of sight intersects the hypothetical galaxy between $l_{min} $ and $l_{max} $}
         \label{drawing}
   \end{figure}

\subsection{The VIMOS far out velocity field}
\label{sec:outflow}
How can we understand the VIMOS results showing a flat velocity curve and an emission 
line width of $\sigma_{los}\sim  80$ \kms\ at galactocentric distances  $\ge 10$ kpc? 
First we note that this value is similar to the average found for a one component fit to 
the nebular gas (and the stars) in the more central regions. 
The spectral resolution of the
VIMOS spectra and the S/N do not allow for a multi component analysis. Below we discuss a few
alternative explanations for the broad lines.

{\bf Thermal broadening:}
Thermal broadening can be ruled out since the implied 
temperature ($\sim10^6$K) would be unrealistic for a gas showing \oiii\ in emission. The 
\oiii\ electron temperature in the centre (where the line is equally broad) is $\sim10^4$K \citep{BO02}.
Hence, the line broadening must be macroscopic  and due to bulk motions. 

{\bf Rotational broadening:}
If the far reaches of \haro\ were a rotating disk, the  line profile would be broadened by 
the fact that the rotational velocity will have different sign and amplitude along 
the line of sight.  Assume we view the galaxy at some impact 
parameter $b$ according to Fig. \ref{drawing} and that the galaxy has a constant rotational
speed $v_{rot}$ which we observe at an inclination $i$. The observed velocity is 
$v_{obs}=v_{rot}  \sin{(i)}$ and from Fig. \ref{vimos} we have that  $v_{obs}= 50 $ \kms .
At $b>>0$ the line  would be made up by components with velocities ranging from 
$\sim v_{obs}/\sqrt{2}\sim 35$\kms\ (at $l=l_{min}$ and $l_{max}$) to $v_{obs}$ at $l=0$,  and the induced 
line broadening would be $\sigma_{obs} \sim (v_{obs} - v_{obs}/\sqrt{2})/2.35 \approx 6 $ \kms 
(where the factor $1/2.35$ comes from the transformation of FWHM to $\sigma$).
 Hence, rotation give a negligible contribution to the line width. 
 
{\bf A radial outflow:}
A scenario which is compatible with a flat velocity gradient and broad emission lines is a uniform 
radial expansion, since this does not affect the velocity
centroid (i.e. the rotation curve) but broadens the line.
A spectrum at a projected galactocentric distance $b$ would 
 sample velocities with line of sight components $v_{los}=v_{exp}\times l/\sqrt{(l^2 + b^2)}$ 
from $l_{min}$ to $l_{max}$, see Fig \ref{drawing}. At $l=0$ the expansion is purely transversal,
whereas $l<0$ and $l>0$ produce  blue-  and red-shifted  components, respectively, which broaden the line. 
The resulting broadening in FWHM would be approximately $v_{exp}$ which hence would 
have to be $\sim200$ \kms\ in order to explain the width of the line.
The line width would be expected to get narrower at larger radii
and there is indeed a weak trend in that direction.

However, a real outflow is unlikely to be spherically symmetric. For instance \citet{Johnson12}
investigated the nearby dwarf galaxy NGC\,1569 in ionised gas and H{\sc i} finding very extended
and irregular bipolar velocity structures. If the far out velocity field in \haro\ is due to outflows, 
it would be  coincidence that the VIMOS FOV happened to capture them.
\citet{Kunth98} found evidence for an outflow (of neutral gas) with 58 km/s from HST/GHRS 
spectroscopy in a pointing towards the centre of \haro .
Additional sight lines were probed by Na absorption in \citet{Sandberg2013} who found an outflow of 44 \kms\ towards \kb\ but an {\em 
inflow} of 27 \kms\ towards \kc . In conclusion, there is no evidence for a galaxy wide radial outflow of the
magnitude needed to explain the halo line width.

{\bf Turbulence from feedback:}
It has been debated whether emission line widths are due to virial motions or due to turbulence 
created through feedback from star formation \citep{TM81,Green10, Moiseev12}. 
The fact that stellar and gas velocity dispersions tend to agree in \haro\ 
 argues against turbulent  feedback driving the gas velocity dispersion.  
 	Moreover, the mechanical energy output of an instantaneous  
starburst peaks at about 20 Myr \citep[Starburst99,][]{Starburst99}. From analysis of the 
luminous star cluster population  \citet{Adamo_haro11} found that the star formation 
rate in \haro\ started to increase rapidly  20 Myr ago. Hence, the current starburst has had
limited time to transport mechanical energy out into the halo. For a  scale of 10 kpc and a 
time scale of 10 Myr, the implied velocity is 1000 km/s. We see no indications of a wind
at this speed, and hence the line width is unlikely to have been produced
purely by feedback. 

{\bf Virial motions:}
We are left with the most plausible explanation being that the line width is mainly due to 
macroscopic bulk motions of the ions. We can then use the clouds as tracers of the 
gravitational potential and equation (1) to get an order of magnitude dynamical mass estimate of
$\sim 10^{11}$\msun. The rotational velocity ($\sim 50/\sin(i)$ \kms at $r=12$kpc) provides another 
$5\times10^9/\sin^2(i)$ \msun 
but is negligible in this context, and the 'halo' of Haro\,11 would  be dominated by
velocity dispersion. 
This assumes dynamical equilibrium and isotropic velocity dispersion, 
both which are  questionable for a merging system like \haro , but the order of magnitude dynamical mass should nevertheless be $10^{11}$\msun
in this scenario. 
 If the merging galaxies came in on parabolic orbits, pre virialisation velocities
will be $\sqrt{2}$ higher and the inferred virial mass a factor of 2 too high.

To conclude, \haro\ is rotating to the largest scales we can probe, and the rotationally
supported mass is, under the assumption of circular motions, $\sim10^{10}$ \msun. 
The emission lines in the halo region are broad, $\sigma \sim 80$ \kms . We find that
the line width is likely due to bulk motions indicating a mass of $\sim10^{11}$ \msun.

\subsection{Stellar  and dynamical masses} 
From multi-wavelength data it is possible to estimate the mass in stars through comparison
with spectral evolutionary synthesis models, under the assumption of a stellar initial mass function (IMF) and a
certain star formation history. This is commonly referred to as {\it stellar mass}. While the resulting age
and star formation history is subject to degeneracies, the stellar mass is often more robust \citep[for reasons discussed in e.g.][]{o01}.

\citet{o01} found that for radii  $r>1 $ kpc, the stellar mass distribution in \haro\  could not be
supported by rotation. Only by assuming that the observed \ha\ velocity dispersion trace mass 
could the dynamical and stellar masses be reconciled. Then the inferred dynamical mass is
$1.9\times10^{10}$ \msun \ for an effective radius of 2.8 kpc and 
$\sigma=81$ \kms .

 In this paper we also make use of the stellar 
population modelling of \haro\ performed in \citet{Hayes07}. In that paper, the stellar population was 
modelled in a spatially resolved fashion using HST multi-band  imaging  from 1500 \AA\ to the I-band,
and  with 2  populations (one young and one old) plus one
ionised gas component. The underlying 
assumptions were that both populations are single stellar populations, calculated with Starburst99 and
Geneva tracks with metallicity $Z=0.004$ and a \citet{Salpeter1955} IMF.  If we ntegrate the stellar mass
from the reference position where the slits cross {\bf ($r=0$)} and out to  $r=10$\arcsec\
we find a total stellar mass of $2.4\times10^{10}$ \msun . Half the quoted mass  is contributed by the inner 
$r<3$\arcsec\ that encompasses the three main knots. The stellar mass estimate is primarily sensitive to 
the assumed low mass slope and cutoff of the IMF.  For more realistic IMFs where the slope flattens a low
mass \citep[e.g.][]{Scalo1986,Chabrier2003} this estimate 
should be reduced to about 60\% of  the quoted value, or $\approx 1.4\times10^{10} $\msun .
These estimates are of the same order as those  by \citet{o01} ($M_\star=1.6\times10^{10}\msun$) 
and \citet{madden2014} ($M_\star=1.7\times10^{10}\msun$), both studies using independent data.

We see a sharp velocity gradient along slit \sca\ from $-6$\arcsec\ to $+4$\arcsec\ (i.e. a total extent of 4 kpc) over
which the velocity varies monotonically  with 200 \kms . Interpreting this in terms of rotation gives a rotationally
supported mass of $4.6\times10^9/\sin^2(i)$ \msun , which for $i>30$\degr\ is smaller than the
mass inferred from the line width (if using the same radius). The inclination is not well constrained, but rotation 
likely gives a significant contribution in  the central ($r \le 10$\arcsec ) region. 
Further out, we can follow the \ha\ kinematics along slit \sca\ from  $-12$\arcsec\ 
to $+16$\arcsec\ (a total scale of 11 kpc). The velocities changes slope on both sides, but the gradients 
are about equally steep, and the same is observed  in the CIGALE velocity field (Fig. \ref{cigale}).

 Hence the velocity field in this ongoing merger is complex. If the velocity gradients are interpreted in terms of 
gravitational motions,  there seems to be ample dynamical mass also outside the central 4 kpc. The velocity 
amplitudes and $\sigma$ are the same as in the centre but the spatial extent almost 3 times larger. The total 
dynamical mass for the \haro\ system may under this assumption approach $10^{11}$\msun . This is close to 
the value inferred from the VIMOS far out velocity field discussed in the previous subsection

\begin{figure*} 
   \centering
   \includegraphics[angle=0,width=1\textwidth]{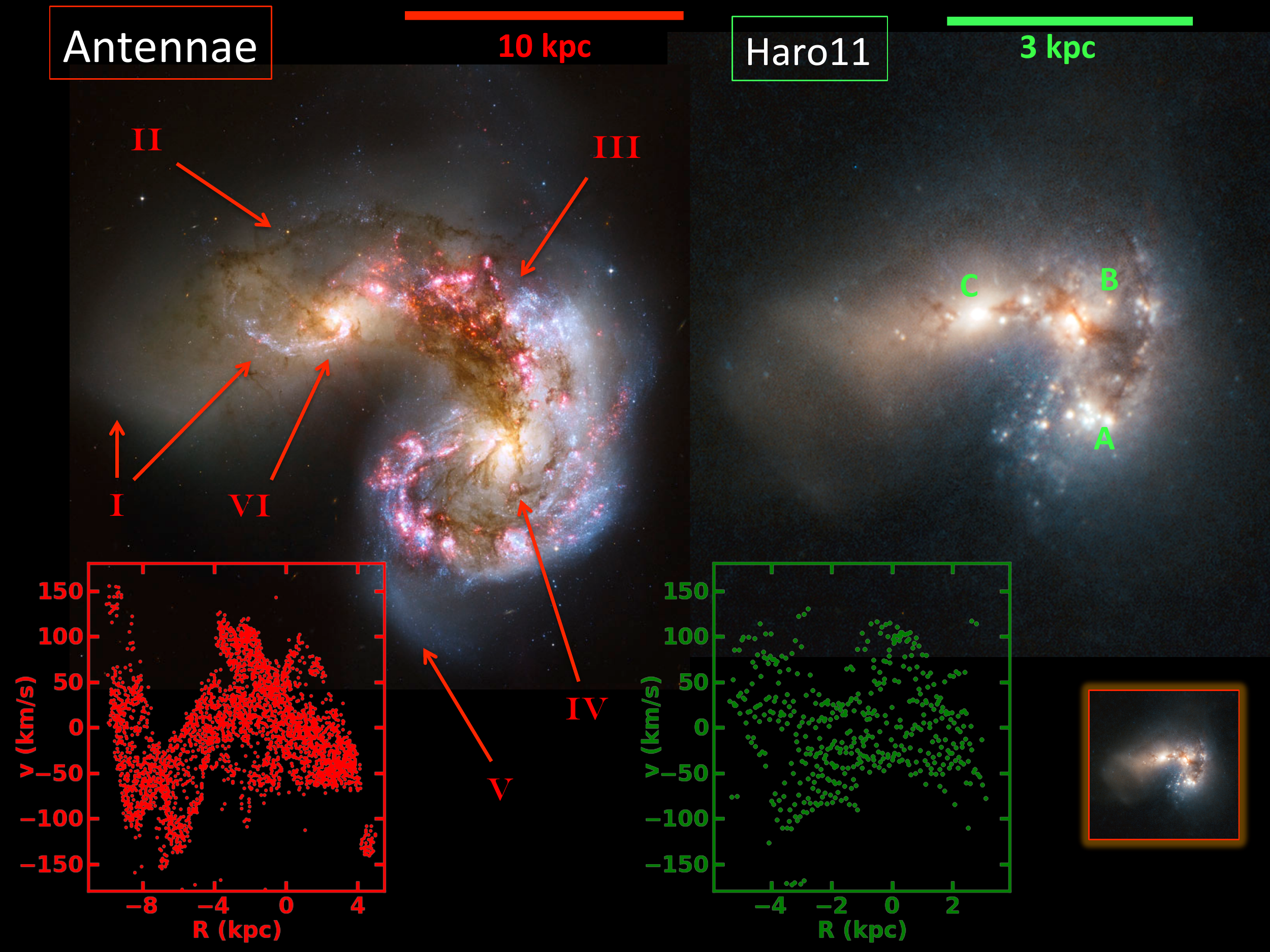}
      \caption{Antennae (left, credit STScI and B. Whitmore) and \haro\ (right, \citet{Adamo_haro11}, credit ESO/ESA/Hubble and NASA) put next to each other.
      The Antennae image has been rotated to the same apparent orientation as \haro\ (which is shown as N up E left)
      and the images have been scaled so that the two systems look equally big on screen, for ease of comparison.
      The physical scale for the image of each system is shown by the red/green bar. 
      In the image of \haro\ we have labelled the three main knots according to their common designation A, B and C (Hayes et al. 2007).
      In the image of the Antenna we point out with arrows and roman numerals some regions of similarity which are discussed in text.
      In the lower right  inset we show the image of \haro\ at the same physical scale of the Antennae.
      The image of \haro\ is a composite of observations from HST/ACS (filters F435W, F550W, F814W) and VLT/NACO  (K$_s$). 
   The inset plot in red (lower left) show the position-velocity (PV) diagram for the Antennae \citep{Amram92} and
   the green inset to the right shows the same for \haro  (this work).
   The PV diagrams have been extracted through the two nuclei in the Antennae and the corresponding
   for \haro\ as seen in Fig. \ref{ant}, i.e. knots \kc\ and \ka . Both PV diagrams cover the same total range
   in velocity ($\Delta v=250$ \kms ). The Antennae VF extend over 130\arcsec (16 kpc) while  the one in 
   \haro\ covers 20\arcsec (8 kpc) and the same velocity scale. As can be seen the shape and amplitude 
   of the PV diagrams   are remarkably similar.}
         \label{ant}
   \end{figure*} 

\label{sec:discussion}

\subsection{Haro\,11 - a denser version of the Antennae}
\label{antcomp}

\haro\ presents a very rich and complex velocity structure. Given the morphological features seen in
the $1.6\mu$m HST/NICMOS image (Fig. \ref{nic3})  this comes as no surprise. However, \haro\ presents yet more intriguing
features.
Comparing it to the Antennae galaxies reveal striking similarities. We demonstrate this in
Fig. \ref{ant} where the HST images of the Antennae\footnote{STScI News Release Number: STScI-2006-46, see:  \\
http://hubblesite.org/newscenter/archive/releases/2006/46} have been rotated and scaled to the same 
apparent orientation and size as \haro . On detailed inspection they almost almost look like copies and 
bright irregular features and dust lanes in  \haro\ have close 
counterparts at the same apparent place in the Antennae and vice versa. 

In Fig. \ref{ant} we highlight, using roman numerals,
some conspicuous features which are also discussed below: 
Both galaxies present distinct edges in the underlying stellar distribution ({\bf I}).
Dust streaks that in the reference system of \haro\ runs horizontally north of knot \kc\ ({\bf II}).
A dusty \ha -bright region at \kb\  ({\bf III}).
The region which in \haro\ is called \ka\ is bright in \ha\ and has many optically bright and blue star clusters SE of it ({\bf IV}).
A faint tidal feature south-east of \ka\ ({\bf V}).
An optically bright compact nucleus \kc\ with dust lanes leading into it ({\bf VI}).
Also the 'ear' in \haro\ may have its analogue in the Antennae system \citep[see e.g. Sect. 8 in ][]{w10}.
Of course, there are differences too: knot \ka\ is a complex of star clusters rather than a nucleus as in the Antennae,
and \haro\ lack extended tidal tails (see below).
Nevertheless, the striking similarity of the the two systems suggest a similar origin - the mergers appear equally advanced
and has presumably been created by progenitor galaxies that are similar to both systems and that has had 
similar orbits with respect to each other. 
However, despite the striking similarities there is one important difference: \haro\ is physically 4 times smaller 
than the Antennae.

The ionised gas velocity field of the Antennae was studied by \citet{Amram92} where they used CIGALE, i.e. the same Fabry-Perot
interferometer as in this work, to study the \ha\ line. The smaller distance and
larger physical size  allow for greater detail to be seen in the  Antennae, so we compare gross properties
(in the reference system of \haro ). The region around \kc\ in both galaxies show low velocity, the 
region around \ka\ have significantly higher velocities, and with \kb\ being intermediate. Hence, not only
the image projected on the sky plane is similar in both galaxies, but also the relative motions.  
\citet{Amram92} showed a position-velocity (PV) diagram, extracted along the line joining the two nuclei, i.e. 
corresponding slit \sca . We extracted a PV diagram from our CIGALE data  cube along this direction and compare it with that of Antennae
in Fig. \ref{ant}. Again, the nearer distance to Antennae allows for more detail and a larger number of
velocity points to be extracted but overall the two PV diagrams are strikingly similar: they share the same 
overall shape and have the same velocity amplitudes ($\Delta v \approx 250$ \kms).
The recent evolution of the SFR in the two galaxies also appears  to be similar. From studying the ages and masses of
young star clusters \citet{Adamo_haro11} find that the SFR in \haro\ has increased with a factor of 50 over the last 40 Myr. 
This corresponds to the estimated time of the 2nd pericenter passage of the merging galaxies in the Antennae after which 
the SFR has increased with a factor of 40  \citep{Karl10,Whitmore2007}. Hence the time scales for the two mergers seems 
to be similar.

While knots \ka\ and \kc\ are separated by 2 kpc, and the nuclei in Antennae by 8 kpc (indicating a factor 
of 4 difference in size) the PV diagram in \haro\ appears relatively more stretched. The distance between the 
extreme velocity values is 3.7 kpc in \haro and 7.6 in Antennae. This is because in \haro\, \kc\ and \ka\ does not 
lie at the extreme values, but the velocity curve continues with the same shape beyond them. This is obvious
also in Figures \ref{cigale} and \ref{f-rc-2}. Hence \haro\ is more compact in position space than in velocity space.
Taking into account that the beam smearing is larger for the more distant \haro\  we determine the velocity gradient 
is $\sim 2.2$ times steeper in \haro .
The dynamical mass  and density of a system scales as:

$${\mathcal M}\propto rv^2 ~~~,~~~\rho \propto {\mathcal M}/r^3 $$


$$\Rightarrow \rho \propto (v/r)^2$$

The morphological similarity suggests that inclinations are similar for \haro\ and the Antennae.
Therefore our results suggest that the mass density in the central parts of \haro\ is 
$2.2^2 \sim5$ times higher than that of the Antennae. Of course, given the non equilibrium nature of mergers and the
uncertainty of inclinations  one should not take masses at face value, but the physical compactness
of \haro\ in any case argues for a higher density as compared to Antennae.

The star formation rate density is known to, on average, scale with the gas surface density as 
$\Sigma_{\rm SFR} \propto \Sigma_{\rm gas}^{1.4}$ \citep{K98}. Since {\bf $\Sigma=\rho r$}
the difference in surface density is again 2.2 and hence $\Sigma_{\rm SFR}$ should be a factor 
of  $2.2^{1.4} = 3$ higher in \haro , if the gas mass fraction is similar in the two systems. 

 The total star formation rate, $SFR = r^2\Sigma_{\rm SFR}$, and we thus expect the total star formation rate 
of the Antennae to be $2.2^2/2.2^{1.4}=1.6$ times  that of  \haro . Hayes et al. (2007) quote a probable SFR 
in the range  20 to 25 \msunyr\ and \citet{madden2014} 
finds 29 for \haro\  whereas \citet{Karl10} compile various estimates for the Antennae in the range 10 to 20 \msunyr . 
The \ha\ luminosity (uncorrected for extinction) is about 3 times higher in \haro\ and the 
IRAS $60 \mu $m luminosity is a factor of 2 higher (based on the values in NED). Hence \haro\ seems to be 
forming stars at 2-3 times the rate of the Antennae, contrary to the expectation from the mass scaling. 

To get a proxy for the stellar mass ratio of the two galaxy systems we consider the K-band luminosities. 
NED gives $K_s=7.2$  for the 
Antennae, or  (for $m-M=31.9$)   $M_K=-24.7$, while  \haro\ has  $M_K=-22.9$ \citep{BO02}. 
Hence the K-band luminosity of the Antennae is   5 times higher than that of \haro , and the stellar mass
likely at least  10 times higher if one accounting for the fact that Antennae is more dusty and that \haro\ due 
 to its  a stronger starburst should have lower $M/L_K$. Hence, the specific star formation rate is at least
 20 times higher in \haro\ than in the Antennae.

The formation of tidal tails in mergers have been investigated in several simulation studies.
Early claims that lack of tails indicated very high dark matter fractions \citep{Dubinski96}
 were countered by \citet{Springel99} who showed that tails could be 
produced for high dark matter fractions provided the halo spin is sufficiently high. Hence the progenitor
galaxies of the \haro\ merger may have had smaller angular momentum or less extended 
dark haloes  than  the Antennae  progenitors, both options suggesting more compact 
participating galaxies in \haro . 
A retrograde merger would also lead to weaker tails, but this explanation is unlikely in the case 
of \haro\ given the close similarity with the Antennae,  which  has been successfully
modelled with a pro-grade encounter \citep[e.g.][]{Karl10}.
 Certainly, dedicated hydrodynamical simulations would be key to better understand \haro .

{Other merging systems  do not possess such obvious apparent similarity as \haro\ and Antennae. 
Arp\,299 has a somewhat similar morphology. It is halfway in size between \haro\ and Antennae 
and has a FIR luminosity 0.4 dex higher than \haro .  Hence the SFR surface densities in \haro\ and Arp\,299 are comparable. 
While there is an abundance of gas in \haro\ detected through various tracers 
  it is remarkably devoid of  cold  {\sc Hi}   \citep{Bergvall00,machattie2014}. 
The SFR per unit mass is very high in \haro \ and  \citet{o99} estimated that the mass of ionised gas in 
\haro\  could be as high as $10^9$\msun . The detection of  [OIII] lines so far out suggest that the galaxy
may be density bounded which  would be consistent with the claimed direct 
detection of Lyman continuum radiation \citep{Bergvall06,Grimes07,Leitet11}.

To conclude, the morphological and kinematical properties of \haro\ and the Antennae are
stunningly similar, width the exception that the physical size of \haro\ is a factor of $\sim 4$ smaller 
and the implied mass density roughly 5 times higher. The star formation rate being  twice as 
high in \haro\ implies that the star formation efficiency  is higher in \haro .

\subsection{On the origin of the line width}
It is known since long that the ionised gas follow a relation between the line width $\sigma$
and the emission line strength of e.g. \hb\ \citep{TM81}. The reason for such a relation
has been discussed over the years, and while its discoverers have argued that it originates in virial
motions, this has been challenged by others \citep{Green10,Moiseev12} who
argue that the relation with the emission line luminosity, hence star formation rate, indicates that
star formation itself -- through feedback -- broadens the line. In the current context, this question
is important as we have used the line width to estimate masses. 
Feedback is of 
course present also in \haro\ and similar luminous BCGs but the question is if feedback can produce lines
with $\sigma\sim80$ \kms , i.e. affecting the bulk motions of the ionised gas rather than just producing
broad low intensity wings? Any matter in a galaxy must respond to the gravitational potential
and  in non rotating 
systems, the line width will always have a component from virial motions. In \haro\ it is 
evident from this work,  that the broad lines may be composed of many discrete velocity components
and what could be wrongly interpreted as a very turbulent ISM is in fact a collection of moderately
broad lines, which it takes high spectral and spatial resolution plus high signal to noise to reveal.
The BCGs that have so far have their stellar motions investigated through Calcium triplet 
observations show that the \hii\ and stellar line widths are remarkably similar on a global scale
\citep[][this work]{o04, m07, c08}. The motions of these predominantly young stars are not affected 
by feedback and they must have been born with this velocity dispersion. 
 Hence, also the ionised gas 
velocity dispersion must be dominated by graviation.

Studies of the evolution of the emission line width $\sigma$ and $v_{rot}/\sigma$  show both to increase
with increasing redshift \citep[e.g.][]{Wisnioski}. This can be intepreted as due to increased gas mass 
fractions, requireing higher $\sigma$ to be stable according to the Toomre criterion \citep{Toomre64}.
 Therefore we must conclude that the \hii\ line width seen in \haro\ and similar systems
to a large extent must be of gravitational origin. Numerical simulations \citep[e.g.][]{Bournaud2013} also 
show that turbulence drives star formation and not vice versa. The lines are not broad because of 
feedback from strong star formation, it is the turbulent medium which (by increasing the pressure 
and the Jeans mass) increases the star formation rate.

\section{Conclusions}
\label{sec:conclus}

We have investigated the kimematics of ionised gas and stars in \haro , a luminous blue compact galaxy.
\haro\ hosts a very strong starburst in terms of specific star formation rate (or equivalently, birthrate $b$-parameter), and is also
one of the most nearby galaxies which matches $z\sim3$ Lyman Break Galaxies in terms of far UV luminosity and surface
brightness.  The kinematical data have been collected with the ESO VLT and 3.6m telescopes. We also make use of imaging
data from the Hubble Space Telescope.

We find that the stellar and ionised gas kinematics to first order agree very well, both showing the same large
scale features and similar velocity dispersions. Hence, wherever we can trace their motions, stars are moving with similar directions and 
amplitude as the ionised gas. This implies that the bulk of the ionised gas
emission arises from gas whose velocities are governed by gravitational motions. 
The irregularities in the velocity field must hence be attributed to real dynamical
disturbances, providing further evidence that \haro\ is a bona fide merger. This interpretation is also supported
by the morphology, exemplified by the deep HST/NICMOS image showing telltale merger signs. The similar
velocity dispersions in stars and gas also suggests that the \ha\ line width is primarily caused by virial motions.
This suggests that emission lines provide
a good probe of the kinematical state of dwarf starbursts, near and far.

When analysing the ionised gas velocity field at higher spectral resolution with integral field spectroscopy we find
several examples of double and triple line profiles. Only with high spectral resolution can these components
be resolved. We find high and low velocity components that make however a small contribution to the total
intensity and which would therefore be missed by low resolution spectroscopy. We used continuity arguments to 
associate the multiple line 
profiles with kinematical components. Notably, we find an elongated, kinematically cold, redshifted arm.

Using an off-centre IFU observation, we can trace the ionised gas (\oiii\ and \hb) kinematics to large radii ($r>10$kpc), 
where we find a flat velocity curve offset from the centre with just 50 \kms . The line width is the same as in the 
centre. We discuss various alternative explanations for the outer velocity field, such as outflows and turbulence, but 
favour an interpretation where
the outer halo is slowly rotating but mainly supported by velocity dispersion, implying a total dynamical mass
of $10^{11}$\msun .

We finally compare \haro\ to the famous Antennae system and find stunning similarities. \haro\ appears to be a
more compact version of the Antenna but otherwise showing very similar central morphology and kinematics. 
The Antennae is 10 times more massive than \haro\ but only form stars at half the rate. One important
difference is that \haro\ lacks extended tidal arms, which may be due to more compact (denser) galaxies 
participating in the merger.

\begin{acknowledgements}
      We thank a number of unnamed colleagues for interesting discussions. G\"O is a Royal Swedish Academy of 
      Science research fellow (supported by a grant from the Knut \& Alice Wallenbergs foundation). This work was 
      supported by the Swedish Research Council and the Swedish National Space Board. This research has made 
      use of the NASA/IPAC Extragalactic Database (NED) which is operated by the Jet Propulsion Laboratory, California 
      Institute of Technology, under contract with the National Aeronautics and Space Administration. 
      This research has made use of NASA's Astrophysics Data System.
\end{acknowledgements}

\bibliographystyle{aa}
\bibliography{haro11cat7}

\begin{thebibliography}{73}
\expandafter\ifx\csname natexlab\endcsname\relax\def\natexlab#1{#1}\fi

\bibitem[{{Adamo} {et~al.}(2011){Adamo}, {{\"O}stlin}, \&
  {Zackrisson}}]{Adamo11}
{Adamo}, A., {{\"O}stlin}, G., \& {Zackrisson}, E. 2011, \mnras, 417, 1904

\bibitem[{{Adamo} {et~al.}(2010){Adamo}, {{\"O}stlin}, {Zackrisson}, {Hayes},
  {Cumming}, \& {Micheva}}]{Adamo_haro11}
{Adamo}, A., {{\"O}stlin}, G., {Zackrisson}, E., {et~al.} 2010, \mnras, 407,
  870

\bibitem[{{Amram} {et~al.}(1991){Amram}, {Boulesteix}, {Georgelin}, {Laval},
  {Le Coarer}, {Marcelin}, \& {Rosado}}]{Amram91}
{Amram}, P., {Boulesteix}, J., {Georgelin}, Y.~M., {et~al.} 1991, The
  Messenger, 64, 44

\bibitem[{{Amram} {et~al.}(1992){Amram}, {Marcelin}, {Boulesteix}, \& {Le
  Coarer}}]{Amram92}
{Amram}, P., {Marcelin}, M., {Boulesteix}, J., \& {Le Coarer}, E. 1992, \aap,
  266, 106

\bibitem[{{Bergvall} {et~al.}(2000){Bergvall}, {Masegosa}, {{\"O}stlin}, \&
  {Cernicharo}}]{Bergvall00}
{Bergvall}, N., {Masegosa}, J., {{\"O}stlin}, G., \& {Cernicharo}, J. 2000,
  \aap, 359, 41

\bibitem[{{Bergvall} \& {{\"O}stlin}(2002)}]{BO02}
{Bergvall}, N. \& {{\"O}stlin}, G. 2002, \aap, 390, 891

\bibitem[{{Bergvall} {et~al.}(2006){Bergvall}, {Zackrisson}, {Andersson},
  {Arnberg}, {Masegosa}, \& {{\"O}stlin}}]{Bergvall06}
{Bergvall}, N., {Zackrisson}, E., {Andersson}, B.-G., {et~al.} 2006, \aap, 448,
  513

\bibitem[{{Blanton} {et~al.}(2003){Blanton}, {Hogg}, {Bahcall}, {Brinkmann},
  {Britton}, {Connolly}, {Csabai}, {Fukugita}, {Loveday}, {Meiksin}, {Munn},
  {Nichol}, {Okamura}, {Quinn}, {Schneider}, {Shimasaku}, {Strauss}, {Tegmark},
  {Vogeley}, \& {Weinberg}}]{Blanton03}
{Blanton}, M.~R., {Hogg}, D.~W., {Bahcall}, N.~A., {et~al.} 2003, \apj, 592,
  819

\bibitem[{{Boulesteix}(1993)}]{Boulesteix93}
{Boulesteix}, J. 1993, Publications de l'Observatoire de Marseille

\bibitem[{{Bournaud} {et~al.}(2013){Bournaud}, {Perret}, {Renaud}, {Dekel},
  {Elmegreen}, {Elmegreen}, {Teyssier}, {Amram}, {Daddi}, {Duc}, {Elbaz},
  {Epinat}, {Gabor}, {Juneau}, {Kraljic}, \& {Le Floch'}}]{Bournaud2013}
{Bournaud}, F., {Perret}, V., {Renaud}, F., {et~al.} 2013, ArXiv e-prints

\bibitem[{{Cappellari} \& {Emsellem}(2004)}]{CE04}
{Cappellari}, M. \& {Emsellem}, E. 2004, \pasp, 116, 138

\bibitem[{{Chabrier}(2003)}]{Chabrier2003}
{Chabrier}, G. 2003, \pasp, 115, 763

\bibitem[{{Cumming} {et~al.}(2008){Cumming}, {Fathi}, {{\"O}stlin}, {Marquart},
  {M{\'a}rquez}, {Masegosa}, {Bergvall}, \& {Amram}}]{c08}
{Cumming}, R.~J., {Fathi}, K., {{\"O}stlin}, G., {et~al.} 2008, \aap, 479, 725

\bibitem[{{Dubinski} {et~al.}(1996){Dubinski}, {Mihos}, \&
  {Hernquist}}]{Dubinski96}
{Dubinski}, J., {Mihos}, J.~C., \& {Hernquist}, L. 1996, \apj, 462, 576

\bibitem[{{Gallego} {et~al.}(1995){Gallego}, {Zamorano}, {Aragon-Salamanca}, \&
  {Rego}}]{Gallego95}
{Gallego}, J., {Zamorano}, J., {Aragon-Salamanca}, A., \& {Rego}, M. 1995,
  \apjl, 455, L1

\bibitem[{{Gil de Paz} {et~al.}(2003){Gil de Paz}, {Madore}, \&
  {Pevunova}}]{Gildepaz03}
{Gil de Paz}, A., {Madore}, B.~F., \& {Pevunova}, O. 2003, \apjs, 147, 29

\bibitem[{{Gon{\c c}alves} {et~al.}(2010){Gon{\c c}alves}, {Basu-Zych},
  {Overzier}, {Martin}, {Law}, {Schiminovich}, {Wyder}, {Mallery}, {Rich}, \&
  {Heckman}}]{Goncalves2010}
{Gon{\c c}alves}, T.~S., {Basu-Zych}, A., {Overzier}, R., {et~al.} 2010, \apj,
  724, 1373

\bibitem[{{Green} {et~al.}(2010){Green}, {Glazebrook}, {McGregor}, {Abraham},
  {Poole}, {Damjanov}, {McCarthy}, {Colless}, \& {Sharp}}]{Green10}
{Green}, A.~W., {Glazebrook}, K., {McGregor}, P.~J., {et~al.} 2010, \nat, 467,
  684

\bibitem[{{Grimes} {et~al.}(2006){Grimes}, {Heckman}, {Hoopes}, {Strickland},
  {Aloisi}, {Meurer}, \& {Ptak}}]{Grimes06}
{Grimes}, J.~P., {Heckman}, T., {Hoopes}, C., {et~al.} 2006, \apj, 648, 310

\bibitem[{{Grimes} {et~al.}(2007){Grimes}, {Heckman}, {Strickland}, {Dixon},
  {Sembach}, {Overzier}, {Hoopes}, {Aloisi}, \& {Ptak}}]{Grimes07}
{Grimes}, J.~P., {Heckman}, T., {Strickland}, D., {et~al.} 2007, \apj, 668, 891

\bibitem[{{Hayes} {et~al.}(2007){Hayes}, {{\"O}stlin}, {Atek}, {Kunth},
  {Mas-Hesse}, {Leitherer}, {Jim{\'e}nez-Bail{\'o}n}, \& {Adamo}}]{Hayes07}
{Hayes}, M., {{\"O}stlin}, G., {Atek}, H., {et~al.} 2007, \mnras, 382, 1465

\bibitem[{{Heckman} {et~al.}(2005){Heckman}, {Hoopes}, {Seibert}, {Martin},
  {Salim}, {Rich}, {Kauffmann}, {Charlot}, {Barlow}, {Bianchi}, {Byun},
  {Donas}, {Forster}, {Friedman}, {Jelinsky}, {Lee}, {Madore}, {Malina},
  {Milliard}, {Morrissey}, {Neff}, {Schiminovich}, {Siegmund}, {Small},
  {Szalay}, {Welsh}, \& {Wyder}}]{Heckman2005}
{Heckman}, T.~M., {Hoopes}, C.~G., {Seibert}, M., {et~al.} 2005, \apjl, 619,
  L35

\bibitem[{{Hippelein} \& {Muench}(1981)}]{HM81}
{Hippelein}, H. \& {Muench}, G. 1981, \aap, 95, 100

\bibitem[{{Hoopes} {et~al.}(2007){Hoopes}, {Heckman}, {Salim}, {Seibert},
  {Tremonti}, {Schiminovich}, {Rich}, {Martin}, {Charlot}, {Kauffmann},
  {Forster}, {Friedman}, {Morrissey}, {Neff}, {Small}, {Wyder}, {Bianchi},
  {Donas}, {Lee}, {Madore}, {Milliard}, {Szalay}, {Welsh}, \& {Yi}}]{Hoopes07}
{Hoopes}, C.~G., {Heckman}, T.~M., {Salim}, S., {et~al.} 2007, \apjs, 173, 441

\bibitem[{{James} {et~al.}(2013){James}, {Tsamis}, {Walsh}, {Barlow}, \&
  {Westmoquette}}]{James2013}
{James}, B.~L., {Tsamis}, Y.~G., {Walsh}, J.~R., {Barlow}, M.~J., \&
  {Westmoquette}, M.~S. 2013, \mnras, 430, 2097

\bibitem[{{Johnson} {et~al.}(2012){Johnson}, {Hunter}, {Oh}, {Zhang},
  {Elmegreen}, {Brinks}, {Tollerud}, \& {Herrmann}}]{Johnson12}
{Johnson}, M., {Hunter}, D.~A., {Oh}, S.-H., {et~al.} 2012, ArXiv e-prints

\bibitem[{{Karl} {et~al.}(2010){Karl}, {Naab}, {Johansson}, {Kotarba}, {Boily},
  {Renaud}, \& {Theis}}]{Karl10}
{Karl}, S.~J., {Naab}, T., {Johansson}, P.~H., {et~al.} 2010, \apjl, 715, L88

\bibitem[{{Kaufer} {et~al.}(2003){Kaufer}, {Pasquini}, {Castillo}, {Schmutzer},
  \& {Smoker}}]{2003Msngr.113...15K}
{Kaufer}, A., {Pasquini}, L., {Castillo}, R., {Schmutzer}, R., \& {Smoker}, J.
  2003, The Messenger, 113, 15

\bibitem[{{Kennicutt}(1998)}]{K98}
{Kennicutt}, Jr., R.~C. 1998, \araa, 36, 189

\bibitem[{{Kobulnicky} \& {Gebhardt}(2000)}]{kg00}
{Kobulnicky}, H.~A. \& {Gebhardt}, K. 2000, \aj, 119, 1608

\bibitem[{{Kochanek} {et~al.}(2001){Kochanek}, {Pahre}, {Falco}, {Huchra},
  {Mader}, {Jarrett}, {Chester}, {Cutri}, \& {Schneider}}]{Kochanek01}
{Kochanek}, C.~S., {Pahre}, M.~A., {Falco}, E.~E., {et~al.} 2001, \apj, 560,
  566

\bibitem[{{Kunth} {et~al.}(1998){Kunth}, {Mas-Hesse}, {Terlevich}, {Terlevich},
  {Lequeux}, \& {Fall}}]{Kunth98}
{Kunth}, D., {Mas-Hesse}, J.~M., {Terlevich}, E., {et~al.} 1998, \aap, 334, 11

\bibitem[{{Kunth} \& {{\"O}stlin}(2000)}]{KO}
{Kunth}, D. \& {{\"O}stlin}, G. 2000, \aapr, 10, 1

\bibitem[{{Larsen}(1999)}]{Larsen99}
{Larsen}, S.~S. 1999, \aaps, 139, 393

\bibitem[{{Leitet} {et~al.}(2011){Leitet}, {Bergvall}, {Piskunov}, \&
  {Andersson}}]{Leitet11}
{Leitet}, E., {Bergvall}, N., {Piskunov}, N., \& {Andersson}, B.-G. 2011, \aap,
  532, A107

\bibitem[{{Leitherer} {et~al.}(1999){Leitherer}, {Schaerer}, {Goldader},
  {Gonz{\'a}lez Delgado}, {Robert}, {Kune}, {de Mello}, {Devost}, \&
  {Heckman}}]{Starburst99}
{Leitherer}, C., {Schaerer}, D., {Goldader}, J.~D., {et~al.} 1999, \apjs, 123,
  3

\bibitem[{{MacHattie} {et~al.}(2014){MacHattie}, {Irwin}, {Madden}, {Cormier},
  \& {R{\'e}my-Ruyer}}]{machattie2014}
{MacHattie}, J.~A., {Irwin}, J.~A., {Madden}, S.~C., {Cormier}, D., \&
  {R{\'e}my-Ruyer}, A. 2014, \mnras, 438, L66

\bibitem[{{Madden} {et~al.}(2014){Madden}, {R{\'e}my-Ruyer}, {Galametz},
  {Cormier}, {Lebouteiller}, {Galliano}, {Hony}, {Bendo}, {Smith}, {Pohlen},
  {Roussel}, {Sauvage}, {Wu}, {Sturm}, {Poglitsch}, {Contursi}, {Doublier},
  {Baes}, {Barlow}, {Boselli}, {Boquien}, {Carlson}, {Ciesla}, {Cooray},
  {Cortese}, {De Looze}, {Irwin}, {Isaak}, {Kamenetzky}, {Karczewski}, {Lu},
  {MacHattie}, {O'Halloran}, {Parkin}, {Rangwala}, {Schirm}, {Schulz},
  {Spinoglio}, {Vaccari}, {Wilson}, \& {Wozniak}}]{madden2014}
{Madden}, S.~C., {R{\'e}my-Ruyer}, A., {Galametz}, M., {et~al.} 2014, \pasp,
  126, 1079

\bibitem[{{Marquart} {et~al.}(2007){Marquart}, {Fathi}, {{\"O}stlin},
  {Bergvall}, {Cumming}, \& {Amram}}]{m07}
{Marquart}, T., {Fathi}, K., {{\"O}stlin}, G., {et~al.} 2007, \aap, 474, L9

\bibitem[{{Micheva} {et~al.}(2012{\natexlab{a}}){Micheva}, {{\"O}stlin},
  {Bergvall}, {Zackrisson}, {Masegosa}, {Marquez}, {Marquart}, \&
  {Durret}}]{Micheva12a}
{Micheva}, G., {{\"O}stlin}, G., {Bergvall}, N., {et~al.} 2012{\natexlab{a}},
  ArXiv e-prints

\bibitem[{{Micheva} {et~al.}(2012{\natexlab{b}}){Micheva}, {{\"O}stlin},
  {Zackrisson}, {Bergvall}, {Marquart}, {Masegosa}, {Marquez}, {Cumming}, \&
  {Durret}}]{Micheva12b}
{Micheva}, G., {{\"O}stlin}, G., {Zackrisson}, E., {et~al.} 2012{\natexlab{b}},
  ArXiv e-prints

\bibitem[{{Micheva} {et~al.}(2010){Micheva}, {Zackrisson}, {{\"O}stlin},
  {Bergvall}, \& {Pursimo}}]{Micheva2010}
{Micheva}, G., {Zackrisson}, E., {{\"O}stlin}, G., {Bergvall}, N., \&
  {Pursimo}, T. 2010, \mnras, 405, 1203

\bibitem[{{Moiseev} \& {Lozinskaya}(2012)}]{Moiseev12}
{Moiseev}, A.~V. \& {Lozinskaya}, T.~A. 2012, \mnras, 423, 1831

\bibitem[{{Monreal-Ibero} {et~al.}(2006){Monreal-Ibero}, {Roth},
  {Sch{\"o}nberner}, {Steffen}, \& {B{\"o}hm}}]{Monreal06}
{Monreal-Ibero}, A., {Roth}, M.~M., {Sch{\"o}nberner}, D., {Steffen}, M., \&
  {B{\"o}hm}, P. 2006, \nar, 50, 426

\bibitem[{{Nagamine} {et~al.}(2004){Nagamine}, {Springel}, {Hernquist}, \&
  {Machacek}}]{Nagamine04}
{Nagamine}, K., {Springel}, V., {Hernquist}, L., \& {Machacek}, M. 2004,
  \mnras, 350, 385

\bibitem[{{Night} {et~al.}(2006){Night}, {Nagamine}, {Springel}, \&
  {Hernquist}}]{Night06}
{Night}, C., {Nagamine}, K., {Springel}, V., \& {Hernquist}, L. 2006, \mnras,
  366, 705

\bibitem[{{Osterbrock} {et~al.}(1996){Osterbrock}, {Fulbright}, {Martel},
  {Keane}, {Trager}, \& {Basri}}]{o96}
{Osterbrock}, D.~E., {Fulbright}, J.~P., {Martel}, A.~R., {et~al.} 1996, \pasp,
  108, 277

\bibitem[{{{\"O}stlin} {et~al.}(2001){{\"O}stlin}, {Amram}, {Bergvall},
  {Masegosa}, {Boulesteix}, \& {M{\'a}rquez}}]{o01}
{{\"O}stlin}, G., {Amram}, P., {Bergvall}, N., {et~al.} 2001, \aap, 374, 800

\bibitem[{{{\"O}stlin} {et~al.}(1999){{\"O}stlin}, {Amram}, {Masegosa},
  {Bergvall}, \& {Boulesteix}}]{o99}
{{\"O}stlin}, G., {Amram}, P., {Masegosa}, J., {Bergvall}, N., \& {Boulesteix},
  J. 1999, \aaps, 137, 419

\bibitem[{{{\"O}stlin} {et~al.}(2004){{\"O}stlin}, {Cumming}, {Amram},
  {Bergvall}, {Kunth}, {M{\'a}rquez}, {Masegosa}, \& {Zackrisson}}]{o04}
{{\"O}stlin}, G., {Cumming}, R.~J., {Amram}, P., {et~al.} 2004, \aap, 419, L43

\bibitem[{{{\"O}stlin} {et~al.}(2007){{\"O}stlin}, {Cumming}, \&
  {Bergvall}}]{o07}
{{\"O}stlin}, G., {Cumming}, R.~J., \& {Bergvall}, N. 2007, \aap, 461, 471

\bibitem[{{{\"O}stlin} {et~al.}(2009){{\"O}stlin}, {Hayes}, {Kunth},
  {Mas-Hesse}, {Leitherer}, {Petrosian}, \& {Atek}}]{o09}
{{\"O}stlin}, G., {Hayes}, M., {Kunth}, D., {et~al.} 2009, \aj, 138, 923

\bibitem[{{Overzier} {et~al.}(2008){Overzier}, {Heckman}, {Kauffmann},
  {Seibert}, {Rich}, {Basu-Zych}, {Lotz}, {Aloisi}, {Charlot}, {Hoopes},
  {Martin}, {Schiminovich}, \& {Madore}}]{Overzier08}
{Overzier}, R.~A., {Heckman}, T.~M., {Kauffmann}, G., {et~al.} 2008, \apj, 677,
  37

\bibitem[{{Overzier} {et~al.}(2009){Overzier}, {Heckman}, {Tremonti}, {Armus},
  {Basu-Zych}, {Goncalves}, {Rich}, {Martin}, {Ptak}, {Schiminovich}, {Ford},
  {Madore}, \& {Seibert}}]{Overzier09}
{Overzier}, R.~A., {Heckman}, T.~M., {Tremonti}, C., {et~al.} 2009, ArXiv
  e-prints

\bibitem[{{Pasquini} {et~al.}(2002){Pasquini}, {Avila}, {Blecha}, {Cacciari},
  {Cayatte}, {Colless}, {Damiani}, {de Propris}, {Dekker}, {di Marcantonio},
  {Farrell}, {Gillingham}, {Guinouard}, {Hammer}, {Kaufer}, {Hill}, {Marteaud},
  {Modigliani}, {Mulas}, {North}, {Popovic}, {Rossetti}, {Royer}, {Santin},
  {Schmutzer}, {Simond}, {Vola}, {Waller}, \& {Zoccali}}]{Pasquini02}
{Pasquini}, L., {Avila}, G., {Blecha}, A., {et~al.} 2002, The Messenger, 110, 1

\bibitem[{{P{\'e}rez-Gallego} {et~al.}(2011){P{\'e}rez-Gallego}, {Guzm{\'a}n},
  {Castillo-Morales}, {Gallego}, {Castander}, {Garland}, {Gruel}, {Pisano}, \&
  {Zamorano}}]{Perezgallego11}
{P{\'e}rez-Gallego}, J., {Guzm{\'a}n}, R., {Castillo-Morales}, A., {et~al.}
  2011, \mnras, 418, 2350

\bibitem[{{Piskunov} \& {Valenti}(2002)}]{2002A&A...385.1095P}
{Piskunov}, N.~E. \& {Valenti}, J.~A. 2002, \aap, 385, 1095

\bibitem[{{Puech} {et~al.}(2006){Puech}, {Hammer}, {Flores}, {{\"O}stlin}, \&
  {Marquart}}]{Puech}
{Puech}, M., {Hammer}, F., {Flores}, H., {{\"O}stlin}, G., \& {Marquart}, T.
  2006, \aap, 455, 119

\bibitem[{{Salpeter}(1955)}]{Salpeter1955}
{Salpeter}, E.~E. 1955, \apj, 121, 161

\bibitem[{{Sandberg} {et~al.}(2013){Sandberg}, {{\"O}stlin}, {Hayes}, {Fathi},
  {Schaerer}, {Mas-Hesse}, \& {Rivera-Thorsen}}]{Sandberg2013}
{Sandberg}, A., {{\"O}stlin}, G., {Hayes}, M., {et~al.} 2013, \aap, 552, A95

\bibitem[{{Scalo}(1986)}]{Scalo1986}
{Scalo}, J.~M. 1986, \fcp, 11, 1

\bibitem[{{Searle} \& {Sargent}(1972)}]{SS72}
{Searle}, L. \& {Sargent}, W.~L.~W. 1972, \apj, 173, 25

\bibitem[{{Springel} \& {White}(1999)}]{Springel99}
{Springel}, V. \& {White}, S.~D.~M. 1999, \mnras, 307, 162

\bibitem[{{Steidel} {et~al.}(1999){Steidel}, {Adelberger}, {Giavalisco},
  {Dickinson}, \& {Pettini}}]{Steidel99}
{Steidel}, C.~C., {Adelberger}, K.~L., {Giavalisco}, M., {Dickinson}, M., \&
  {Pettini}, M. 1999, \apj, 519, 1

\bibitem[{{Terlevich} \& {Melnick}(1981)}]{TM81}
{Terlevich}, R. \& {Melnick}, J. 1981, \mnras, 195, 839

\bibitem[{{Tonry} \& {Davis}(1979)}]{TD}
{Tonry}, J. \& {Davis}, M. 1979, \aj, 84, 1511

\bibitem[{{Toomre}(1964)}]{Toomre64}
{Toomre}, A. 1964, \apj, 139, 1217

\bibitem[{{Vader} {et~al.}(1993){Vader}, {Frogel}, {Terndrup}, \&
  {Heisler}}]{Vader93}
{Vader}, J.~P., {Frogel}, J.~A., {Terndrup}, D.~M., \& {Heisler}, C.~A. 1993,
  \aj, 106, 1743

\bibitem[{{van Hoof}(1999)}]{AtomicLineList}
{van Hoof}, P. 1999, Atomic Line List, version 2.04

\bibitem[{{Whitmore} {et~al.}(2007){Whitmore}, {Chandar}, \&
  {Fall}}]{Whitmore2007}
{Whitmore}, B.~C., {Chandar}, R., \& {Fall}, S.~M. 2007, \aj, 133, 1067

\bibitem[{{Whitmore} {et~al.}(2010){Whitmore}, {Chandar}, {Schweizer},
  {Rothberg}, {Leitherer}, {Rieke}, {Rieke}, {Blair}, {Mengel}, \&
  {Alonso-Herrero}}]{w10}
{Whitmore}, B.~C., {Chandar}, R., {Schweizer}, F., {et~al.} 2010, \aj, 140, 75

\bibitem[{{Wisnioski} {et~al.}(2015){Wisnioski}, {F{\"o}rster Schreiber},
  {Wuyts}, {Wuyts}, {Bandara}, {Wilman}, {Genzel}, {Bender}, {Davies},
  {Fossati}, {Lang}, {Mendel}, {Beifiori}, {Brammer}, {Chan}, {Fabricius},
  {Fudamoto}, {Kulkarni}, {Kurk}, {Lutz}, {Nelson}, {Momcheva}, {Rosario},
  {Saglia}, {Seitz}, {Tacconi}, \& {van Dokkum}}]{Wisnioski}
{Wisnioski}, E., {F{\"o}rster Schreiber}, N.~M., {Wuyts}, S., {et~al.} 2015,
  \apj, 799, 209

\bibitem[{{Wyder} {et~al.}(2005){Wyder}, {Treyer}, {Milliard}, {Schiminovich},
  {Arnouts}, {Budav{\'a}ri}, {Barlow}, {Bianchi}, {Byun}, {Donas}, {Forster},
  {Friedman}, {Heckman}, {Jelinsky}, {Lee}, {Madore}, {Malina}, {Martin},
  {Morrissey}, {Neff}, {Rich}, {Siegmund}, {Small}, {Szalay}, \&
  {Welsh}}]{Wyder05}
{Wyder}, T.~K., {Treyer}, M.~A., {Milliard}, B., {et~al.} 2005, \apjl, 619, L15

\end{thebibliography}

\end{document}